\documentclass[preprint,12pt,a4paper]{elsarticle}

\usepackage{amsmath,amssymb,amsfonts}
\usepackage{bm}
\usepackage{booktabs}
\usepackage{multirow}
\usepackage{graphicx}
\usepackage{algorithm}
\usepackage{algpseudocode}
\usepackage{enumitem}
\usepackage{xcolor}
\usepackage{microtype}
\usepackage{url}
\usepackage{hyperref}
\usepackage{cleveref}
\usepackage{siunitx}
\usepackage{array}
\usepackage{tabularx}
\usepackage{makecell}
\usepackage{threeparttable}
\usepackage{graphicx}
\usepackage{subcaption}

\journal{Applied Soft Computing}

\definecolor{draftred}{RGB}{170,20,20}

\newcommand{\method}{Agentic LinkDreamer}
\newcommand{\dataset}{V2X-Real}

\newcommand{\stateset}{\mathcal{S}}

\begin{document}

\begin{frontmatter}

\title{Agentic LinkDreamer: Adaptive Neuro-Fuzzy World-Model Decision Making for Intermittent V2X Cooperative Perception}

\author[1]{Zhaoyu Liu\fnref{equal}}
\author[1]{Dongmin Wang\fnref{equal}}
\author[1]{Liu Cao\corref{cor1}}
\ead{liu.cao@cityu-dg.edu.cn}

\author[2]{Jijing Cai}
\author[2]{Kai Fang}

\author[3]{Syed~Ishtiaque~Ahmed}

\author[4]{Almetwally~M.~Mostafa}

\affiliation[1]{
    organization={Electrical and Electronic Engineering Program,
    City University of Hong Kong (Dongguan)},
    city={Dongguan},
    postcode={523808},
    country={China}
}

\affiliation[2]{
    organization={College of Artificial Intelligence,
    Zhejiang A\&F University},
    city={Hangzhou},
    postcode={311300},
    country={China}
}

\affiliation[3]{
    organization={Department of Computer Science,
    University of Toronto},
    city={Toronto},
    country={Canada}
}

\affiliation[4]{
    organization={Information Systems Department,
    College of Computer and Information Sciences,
    King Saud University},
    city={Riyadh},
    postcode={11543},
    country={Saudi Arabia}
}

\fntext[equal]{These authors contributed equally to this work.}

\cortext[cor1]{Corresponding author.}

\begin{abstract}
Intermittent vehicle-to-everything (V2X) communication can abruptly remove collaborator information required by cooperative perception. Existing recovery methods can reconstruct missing features, but typically do not determine when prediction should be trusted, when historical information should be reused, or when communication should be refreshed. This paper proposes \method{}, an adaptive soft-computing framework that integrates a latent world model
with an interpretable neuro-fuzzy decision agent. The framework maintains causal collaborator memory and employs a relative-motion-conditioned recurrent world model to predict unavailable bird's-eye-view (BEV) features during communication outages. Prediction uncertainty, perceptual complementarity, memory freshness, communication quality, and residual communication budget form a compact decision state. A differentiable Takagi--Sugeno controller adaptively selects among ego-only perception, motion-compensated historical
feature reuse, latent imagination, low-rate temporal-summary refresh, and high-rate collaborator-feature refresh. The policy balances detection risk, communication overhead, latency, predictive uncertainty, and action switching,
while reflection after link recovery updates uncertainty calibration and causal memory. To enable reproducible evaluation without assuming unavailable physical-layer traces, we construct a geometry-derived V2X degradation protocol from the recorded trajectories and LiDAR-derived scene geometry of \dataset{}. Experiments on the real-world \dataset{} V2V split show that \method{} achieves
90.95 AP@0.5 on outage frames and recovers 92.9\% of the cooperative-perception performance gap. At the high-performance operating point, it remains only 0.65 AP below the cooperative upper reference while reducing communication
volume by approximately 31\% relative to continuous high-rate cooperation. These results demonstrate that adaptive world-model reasoning and neuro-fuzzy decision making provide a more reliable alternative to unconditional missing-feature reconstruction under intermittent V2X communication.
\end{abstract}




\end{frontmatter}

\section{Introduction}
\label{sec:introduction}

Cooperative perception enables connected vehicles and roadside infrastructure to exchange complementary observations, thereby reducing blind spots, extending sensing range, and improving three-dimensional object detection. Intermediate fusion is particularly attractive because it exchanges semantically informative features without transmitting full raw point clouds~\cite{wang2020v2vnet,xu2022v2xvit,hu2022where2comm,liu2026phylayermodelingthroughputdrivenadaptation,cao2025lightweightcoordinateconditioneddiffusionapproach}. Its effectiveness, however, depends strongly on communication availability. When a collaborator becomes temporarily unavailable because of blockage, distance, scheduling, or packet loss, the ego vehicle may abruptly lose objects and regions that were visible only from the remote viewpoint. Cooperative perception under intermittent V2X communication is therefore not simply a communication problem, but a partially observed multi-agent perception problem in which previously received collaborator information may remain useful.

Existing approaches can mitigate this degradation by reusing historical features, compensating temporal misalignment, or predicting missing collaborator representations. However, feature recovery alone does not determine whether the recovered information should actually be used. During an interruption, a perception system must decide whether to rely on ego-only perception, reuse a stale but real collaborator feature, or trust a predicted representation. When communication becomes available, it must further determine whether the expected perception benefit justifies refreshing compact temporal memory or transmitting a higher-fidelity collaborator representation. These choices depend jointly on communication quality, memory freshness, outage duration, prediction uncertainty, collaborator usefulness, and available communication resources. Consequently, communication-resilient cooperative perception requires an adaptive decision mechanism rather than a fixed recovery strategy.

To address this problem, we propose \method, an adaptive neuro-fuzzy world-model framework that formulates intermittent cooperative perception as an agentic communication--perception decision process. The framework maintains a causal temporal summary together with the latest available high-fidelity collaborator representation. When the current collaborator observation is unavailable, a relative-motion-conditioned latent world model predicts the missing BEV feature from the stored temporal context, the ego vehicle's current observation, and estimated inter-agent motion. Prediction uncertainty, perceptual complementarity, and action-dependent perception risk provide compact decision-relevant information. An interpretable Takagi--Sugeno neuro-fuzzy controller then selects among five operating modes: ego-only perception, historical-feature holding and motion compensation, latent imagination, low-rate temporal-summary refresh, and high-rate collaborator-feature refresh. Feasibility masking prevents unavailable, latency-violating, or unreliable actions from being selected.

The resulting policy balances perception quality against transmitted bits, end-to-end latency, prediction uncertainty, and excessive action switching. Importantly, latent prediction is not applied unconditionally: the agent can reject an uncertain imagined feature or switch to a safer information source according to the current operating state. When a new real collaborator feature becomes available after an interruption, the discrepancy between the predicted and observed representations provides a reflection signal for updating uncertainty calibration and collaborator memory. The resulting procedure forms an observe--imagine--decide--act--reflect loop that integrates prediction, communication, and perception within a unified decision framework.

We evaluate \method{} on \dataset, a real-world multi-agent cooperative-perception benchmark with synchronized sensing and agent-specific poses~\cite{xiang2024v2xreal}. The main experiments use the vehicle-to-vehicle split so that both the ego vehicle and collaborator are moving, which directly supports evaluation of relative-motion-conditioned prediction. Since \dataset{} does not provide synchronized physical-layer channel measurements for every perception frame, we do not interpret its trajectories as radio-accurate communication traces. Instead, inter-agent distance and geometric blockage are used to construct a reproducible, severity-controlled communication-degradation protocol that generates causal V2X outage episodes for evaluation.

The main contributions are summarized as follows.
\begin{enumerate}[leftmargin=1.6em]
\item We formulate cooperative perception under intermittent V2X communication as a partially observed agentic decision problem, in which the ego vehicle explicitly selects among recovery, fallback, and communication-refresh actions instead of applying missing-feature prediction unconditionally.

\item We develop a causal temporal-memory and relative-motion-conditioned latent world model that predicts unavailable collaborator BEV representations and provides decision-relevant uncertainty, perceptual complementarity, and action-dependent perception risk.

\item We design an adaptive and interpretable Takagi--Sugeno neuro-fuzzy controller that selects among ego-only perception, historical-feature reuse, latent imagination, low-rate summary refresh, and high-rate feature refresh while accounting for communication cost, latency, prediction reliability, and action stability.

\item We establish a reproducible geometry-derived V2X interruption protocol on the real-world \dataset{} V2V split and evaluate the proposed framework through outage robustness, communication--perception trade-offs, ablation studies, policy interpretability, and qualitative failure analysis.

\end{enumerate}

The remainder of this paper is organized as follows. Section~2 reviews cooperative perception under imperfect V2X communication, world models and agentic decision making, and neuro-fuzzy decision systems. Section~3 presents the system model and formulates intermittent cooperative perception as a constrained, partially observed multi-objective decision problem. Section~4 details the proposed \method{} framework and its training strategy. Section~5 presents the experimental setup and evaluation results. Finally, Section~6 concludes the paper.

\section{Related work}
\label{sec:related}

\subsection{Cooperative Perception under Imperfect V2X Communication}

Cooperative perception enables connected vehicles and infrastructure to exchange complementary observations and improve perception beyond the capability of a single agent. V2VNet introduced learned message passing for vehicle cooperation~\cite{wang2020v2vnet}, while DiscoNet improved the accuracy--bandwidth trade-off through a distilled collaboration graph~\cite{li2021disconet}. V2X-ViT modeled heterogeneous vehicle--infrastructure interactions using transformer attention~\cite{xu2022v2xvit}, and Where2comm transmitted only perceptually informative regions using spatial confidence maps~\cite{hu2022where2comm}. CoBEVT extended cooperative transformers to BEV perception~\cite{xu2023cobevt}, whereas How2comm jointly considered communication efficiency, delay, and collaboration heterogeneity~\cite{yang2023how2comm}. These studies demonstrate the effectiveness of adaptive information exchange, but generally rely on valid or recoverable collaborator information during cooperative operation.

Communication delay and interruption further challenge cooperative perception. SyncNet compensates asynchronous information through temporal feature prediction~\cite{lei2022syncnet}, while V2X-INCOP reconstructs missing messages using spatial-temporal prediction, knowledge distillation, and curriculum learning~\cite{ren2024incop}. How2comm incorporates flow-guided delay compensation~\cite{yang2023how2comm}, and CoEF exploits entropy and feature reprojection to retain useful information under communication degradation~\cite{li2025coef}. These methods show that historical features can mitigate communication failures. However, recovery is typically based on a predefined compensation mechanism, without explicitly deciding whether a stale real feature, a predicted representation, or a new communication refresh is preferable under the current memory age, uncertainty, and resource condition.

Public benchmarks include simulated datasets such as OPV2V~\cite{xu2022opv2v} and V2X-Sim~\cite{li2022v2xsim}, as well as real-world platforms such as DAIR-V2X~\cite{yu2022dairv2x} and V2X-Real~\cite{xiang2024v2xreal}. We adopt the V2X-Real V2V setting because its moving-agent trajectories and agent-specific poses support temporal collaborator modeling and relative-motion-conditioned evaluation under intermittent communication.

\subsection{World Models and Agentic Decision Making}

World models learn compact latent dynamics that support prediction, planning, and decision making. Early world models combined learned visual representations, recurrent dynamics, and a controller~\cite{ha2018worldmodels}. Dreamer optimized behavior through imagined latent trajectories~\cite{hafner2020dreamer}, while MuZero demonstrated planning with learned latent dynamics without explicitly reconstructing raw observations~\cite{schrittwieser2020muzero}. The Dyna framework similarly connected real experience, model learning, and simulated decision updates~\cite{sutton1991dyna}. These approaches establish latent prediction as an effective mechanism for reasoning about states that are not directly observed.

This idea is particularly relevant to intermittent cooperative perception, where the current collaborator observation may be unavailable while historical information remains informative. Nevertheless, predicting a missing representation alone does not determine whether it should be trusted or used by the detector. In Agentic LinkDreamer, the world model therefore serves as a predictive component within an agentic decision loop. It estimates the missing collaborator BEV representation from causal temporal memory, the ego observation, and relative motion, while prediction uncertainty and action-dependent perception risk provide decision-relevant information. The agent can consequently choose among ego-only perception, historical-feature reuse, latent imagination, and communication refresh instead of applying prediction unconditionally.

\subsection{Neuro-Fuzzy Decision Systems and Research Positioning}

Soft computing provides useful mechanisms for reasoning with continuous and uncertain operating states. Fuzzy sets represent linguistic concepts using graded memberships~\cite{zadeh1965fuzzy}, while fuzzy decision theory supports decision making under uncertain objectives and constraints~\cite{bellman1970decision}. Takagi--Sugeno systems combine fuzzy antecedents with smooth functional consequents~\cite{takagi1985fuzzy}, and adaptive neuro-fuzzy inference systems enable membership and rule parameters to be learned from data~\cite{jang1993anfis}. Such properties are suitable for V2X decision variables including communication quality, memory age, outage depth, prediction uncertainty, perceptual complementarity, and residual communication budget, for which universal crisp thresholds are difficult to define.

A purely neural decision policy can learn nonlinear mappings from these variables to communication actions, but its decision mechanism is difficult to interpret. Conversely, manually designed fuzzy rules provide transparency but may not adapt well to perception statistics and changing communication conditions. We therefore employ a differentiable Takagi--Sugeno controller whose membership functions, rule importance, and action consequents are learned jointly while retaining physically meaningful decision variables.

Overall, existing interruption-aware cooperative perception mainly focuses on recovering unavailable features, while predictive world models mainly provide latent forecasting and conventional fuzzy systems mainly address uncertain decision boundaries. These directions have rarely been integrated to jointly reason about temporal memory, prediction reliability, collaborator usefulness, communication cost, and action feasibility. To address this gap, we formulate intermittent V2X cooperative perception as an agentic decision problem in which the ego vehicle adaptively selects among ego-only perception, historical-feature holding, latent imagination, low-rate summary refresh, and high-rate feature refresh according to the current perception and communication state.

\section{System Model and Problem Formulation}
\label{sec:problem}
\label{sec:system_model}

\begin{figure*}[t]
    \centering
    \includegraphics[width=1\textwidth]{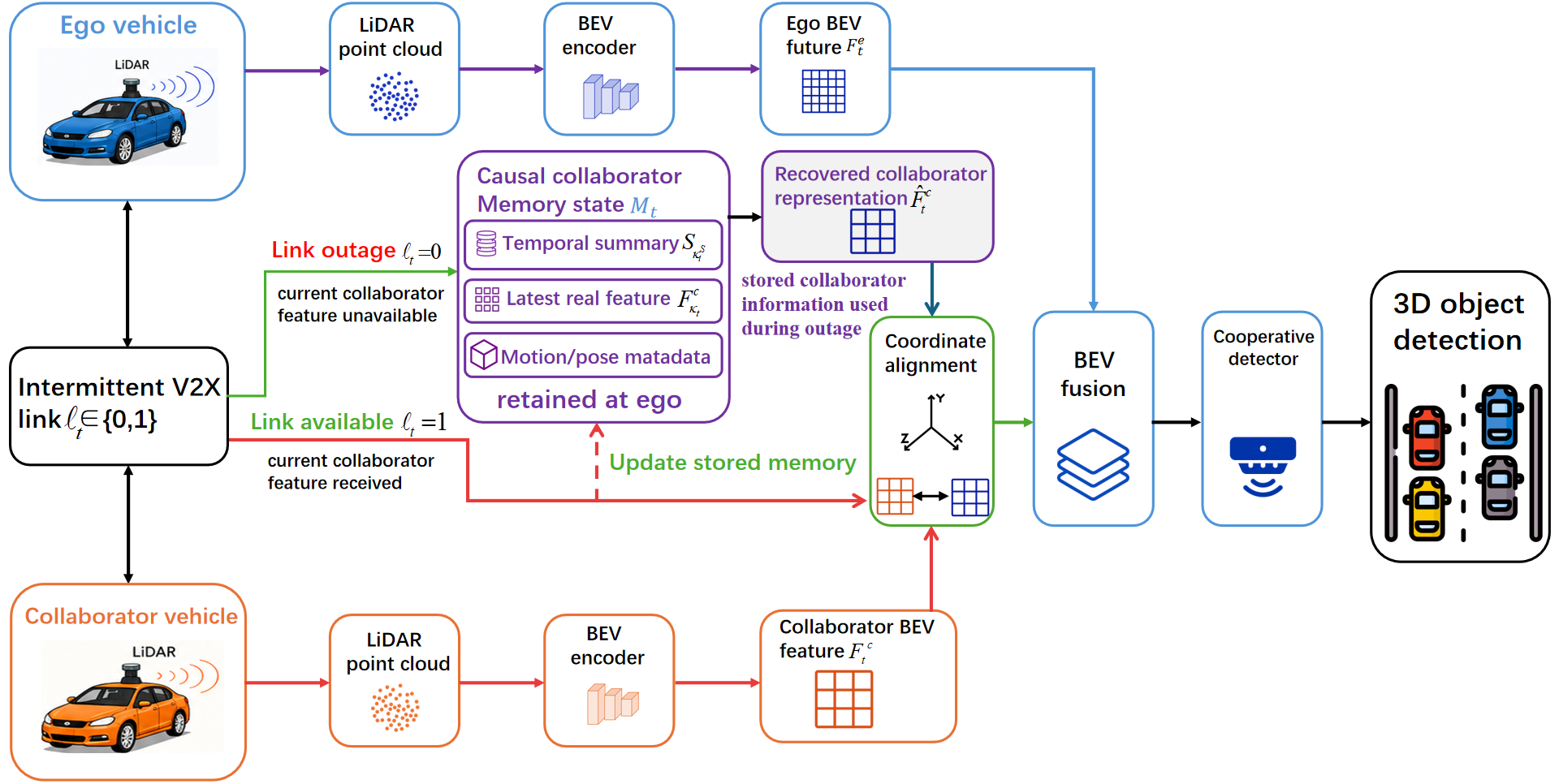}
    \caption{System model of cooperative perception under intermittent V2X communication.}
    \label{fig:system_model}
\end{figure*}

Consider the V2X-enabled cooperative perception system shown in Fig.~\ref{fig:system_model}, consisting of an ego vehicle $e$ and a collaborator $c$. Both vehicles are equipped with LiDAR sensors and onboard perception processors. At each sensing frame $t\in\{1,2,\ldots\}$, the two vehicles observe the surrounding environment from different viewpoints. When the V2X link is available, the collaborator can transmit an intermediate perception representation to the ego vehicle for cooperative object detection. When communication is temporarily interrupted because of blockage, mobility, packet loss, or network congestion, the current collaborator observation becomes unavailable.

Instead of assuming continuous feature transmission, the ego vehicle selects among different communication--perception operating modes according to the current link condition, previously received information, prediction reliability, and expected cooperative benefit. This section first models cooperative perception under intermittent V2X communication, then defines the compact agent state and feasible action space, and finally formulates the resulting sequential decision problem.

\subsection{Cooperative Perception under Intermittent V2X}
\label{sec:partial_observability}

Let $X_t^i$ denote the LiDAR point cloud observed by agent $i\in\{e,c\}$ at frame $t$. A shared BEV perception encoder $E_{\mathrm{bev}}(\cdot)$ maps the point cloud into an intermediate feature representation
\begin{equation}
F_t^i= E_{\mathrm{bev}}(X_t^i),
\label{eq:bev_feature}
\end{equation}
where $F_t^i\in\mathbb{R}^{C\times H\times W}$,$C$, $H$, and $W$ denote the number of feature channels and the
height and width of the BEV feature map, respectively.

Because the ego vehicle and collaborator observe the scene in different coordinate systems, the collaborator representation must be transformed into the current ego coordinate frame before fusion. Let $\mathbf{T}_t^e$ and $\mathbf{T}_t^c$ denote the corresponding poses.
The relative transformation from the collaborator frame to the ego frame is
\begin{equation}
\mathbf{T}_{c\rightarrow e,t}
=
(\mathbf{T}_t^e)^{-1}\mathbf{T}_t^c.
\label{eq:relative_transform}
\end{equation}
Accordingly, the spatially aligned collaborator feature is
\begin{equation}
\bar{F}_t^c
=
\mathcal{T}
\left(
F_t^c;
\mathbf{T}_{c\rightarrow e,t}
\right),
\label{eq:aligned_feature}
\end{equation}
where $\mathcal{T}(\cdot)$ denotes a differentiable spatial
transformation from the collaborator frame to the ego frame.

When the current collaborator feature is successfully received, the cooperative detector produces
\begin{equation}
\widehat{Y}_t^{\mathrm{coop}}
=
G
\left(
F_t^e,
\bar{F}_t^c
\right),
\label{eq:cooperative_detection}
\end{equation}
where $G(\cdot,\cdot)$ comprises the intermediate feature-fusion module and the downstream detection head. The output
$\widehat{Y}_t^{\mathrm{coop}}$ contains the predicted object classes, bounding boxes, and confidence scores. For comparison, ego-only perception is expressed as
\begin{equation}
\widehat{Y}_t^{\mathrm{ego}}
=
G
\left(
F_t^e,
\mathbf{0}
\right),
\label{eq:ego_detection}
\end{equation}
where $\mathbf{0}\in\mathbb{R}^{C\times H\times W}$ is an all-zero tensor with the same dimension as the aligned collaborator feature.

The instantaneous V2X-link state is represented by
\begin{equation}
\ell_t
=
\begin{cases}
1, & \text{if a new collaborator message can be received},\\
0, & \text{otherwise}.
\end{cases}
\label{eq:link_state}
\end{equation}
where the variable $\ell_t$ represents the effective communication outcome available to the perception agent and can be obtained from packet delivery information, physical-layer measurements, or a communication-state estimator. In the experiments of this work, it is instantiated using the reproducible V2X degradation protocol.

To preserve causal information across frames, let $\kappa_t$ denote the latest frame at or before $t$ at which a high-fidelity collaborator representation was successfully received, and let $\kappa_t^S$ denote the latest frame at which a valid compact temporal summary was received. The memory available to the ego vehicle is represented by
\begin{equation}
M_t
=
\left\{
S_{\kappa_t^S},
F_{\kappa_t}^{c},
\mathbf{T}_{\kappa_t}^{c},
\mathbf{v}_{\kappa_t}^{c},
\bar{e}_{\kappa_t}
\right\},
\label{eq:memory_state}
\end{equation}
where $S_{\kappa_t^S}$ is the latest received temporal summary, $F_{\kappa_t}^{c}$ is the latest high-fidelity collaborator BEV feature, $\mathbf{T}_{\kappa_t}^{c}$ and
$\mathbf{v}_{\kappa_t}^{c}$ denote its associated pose and motion information, respectively, and $\bar{e}_{\kappa_t}$ is the latest reflection-based prediction-calibration statistic. The construction and update of these memory components are described in Section~\ref{sec:method}.

When $\ell_t=0$, the current collaborator feature $F_t^c$ is not observable. Therefore, the ego vehicle makes decisions from
\begin{equation}
\mathcal{O}_t
=
\left\{
F_t^e,
M_t,
q_t,
b_t,
\ell_t
\right\},
\label{eq:observation}
\end{equation}
where $q_t\in[0,1]$ is a normalized indicator of the current
communication quality and $b_t$ denotes the instantaneous residual communication payload budget available at frame $t$. Hence, under intermittent V2X communication, the current remote observation may be hidden from the ego agent, leading to a partially observed sequential decision problem.

\subsection{Agent State and Feasible Action Space}
\label{sec:state_action}
\label{sec:agent_state}

Directly using the high-dimensional observation $\mathcal{O}_t$ for online decision making is inefficient. We construct a compact state that retains communication status, memory freshness, and decision-relevant perception information.

The short-term change in communication quality is defined as
\begin{equation}
\Delta q_t
=
q_t-q_{t-1},
\label{eq:quality_trend}
\end{equation}
which distinguishes improving and degrading link conditions.

Two memory-age variables are introduced because low-rate temporal summaries and high-fidelity collaborator features can be refreshed at different times. The temporal-summary age is
\begin{equation}
a_t^S
=
t-\kappa_t^S,
\label{eq:summary_age}
\end{equation}
then the age of the latest high-fidelity collaborator feature is
\begin{equation}
a_t^H
=
t-\kappa_t.
\label{eq:feature_age}
\end{equation}
The former characterizes the freshness of the temporal context used by the latent world model, while the latter measures the staleness of the feature available to the historical-feature reuse action.

The consecutive communication-outage depth is updated as
\begin{equation}
\tau_t=
\begin{cases}
0, & \ell_t=1,\\
\tau_{t-1}+1, & \ell_t=0,
\end{cases}
\label{eq:outage_depth}
\end{equation}
where $\tau_t$ measures how many consecutive frames have elapsed without a newly received collaborator message.

We further introduce two perception-related decision variables. Specifically, $u_t\in[0,1]$ denotes the predictive uncertainty of the world-model-generated collaborator representation, while $g_t\in[0,1]$ represents perceptual complementarity between the collaborator and ego observations. A large $u_t$ indicates that latent
imagination is unreliable, whereas a large $g_t$ indicates that the collaborator is expected to provide information that cannot be adequately recovered from the ego observation alone. Their estimation is presented in Section~\ref{sec:method}.

The continuous decision descriptor is therefore defined as
\begin{equation}
\widetilde{\bm{s}}_t
=
\left[
q_t,\,
\Delta q_t,\,
a_t^S,\,
a_t^H,\,
\tau_t,\,
u_t,\,
g_t,\,
b_t
\right]^{\top},
\label{eq:continuous_state}
\end{equation}
and the complete decision state is
\begin{equation}
\bm{s}_t
=
\left[
\widetilde{\bm{s}}_t^{\top},
\ell_t
\right]^{\top}
\in\stateset,
\label{eq:state}
\end{equation}
where $\stateset$ denotes the compact decision-state space. All continuous variables in $\widetilde{\bm{s}}_t$ are normalized before being provided to the decision controller. In the proposed implementation, the eight continuous variables are used by the neuro-fuzzy controller, while the binary link state $\ell_t$ is used to enforce the action-feasibility mask.

At each frame, the ego agent selects one action from
\begin{equation}
\mathcal{A}
=
\left\{
a^{\mathrm{ego}},
a^{\mathrm{hold}},
a^{\mathrm{imag}},
a^{\mathrm{refL}},
a^{\mathrm{refH}}
\right\}.
\label{eq:actions}
\end{equation}
where $a^{\mathrm{ego}}$ performs ego-only perception,
$a^{\mathrm{hold}}$ reuses the latest received high-fidelity
collaborator representation, and $a^{\mathrm{imag}}$ predicts the current collaborator representation using the latent world model. The communication actions $a^{\mathrm{refL}}$ and
$a^{\mathrm{refH}}$ request a low-rate compact temporal summary and a high-rate collaborator representation, respectively.

Let $B_L$ and $B_H$ denote the corresponding communication payloads, with $B_H>B_L$. The number of transmitted bits associated with action
$a_t$ is
\begin{equation}
B(a_t)=
\begin{cases}
0,
&
a_t\in
\{a^{\mathrm{ego}},a^{\mathrm{hold}},a^{\mathrm{imag}}\},
\\
B_L,
&
a_t=a^{\mathrm{refL}},
\\
B_H,
&
a_t=a^{\mathrm{refH}}.
\end{cases}
\label{eq:action_bits}
\end{equation}

A communication refresh can only be executed when the current V2X link is available and sufficient instantaneous communication budget remains. Accordingly, the feasible action set is
\begin{equation}
\begin{aligned}
\mathcal{A}_{\mathrm{feas}}(\bm{s}_t)
=&
\{
a^{\mathrm{ego}},
a^{\mathrm{hold}},
a^{\mathrm{imag}}
\}
\\
&\cup
\left\{
a^{\mathrm{refL}}:
\ell_t=1,\,
B_L\leq b_t
\right\}
\\
&\cup
\left\{
a^{\mathrm{refH}}:
\ell_t=1,\,
B_H\leq b_t
\right\}.
\end{aligned}
\label{eq:feasible_actions}
\end{equation}

Equation~\eqref{eq:feasible_actions} enforces causal decision making. In particular, after the link has already become unavailable, the ego agent cannot retroactively request additional information from the collaborator and is therefore restricted to ego-only perception, historical feature reuse, or latent imagination.

\subsection{Multi-Objective Decision Problem}
\label{sec:cost_model}
\label{sec:problem_formulation}

Given $\bm{s}_t$ and $\mathcal{A}_{\mathrm{feas}}(\bm{s}_t)$, the ego agent must select an operating mode that balances perception reliability against communication and latency costs. The candidate actions differ in their expected detection accuracy, communication overhead, processing
latency, prediction reliability, and temporal stability. We define an action-dependent multi-objective stage cost.

Let $\mathcal{L}_{\mathrm{det},t}(a)$ denote the downstream detection loss that would result from executing candidate action $a\in\mathcal{A}_{\mathrm{feas}}(\bm{s}_t)$. Since its ground-truth value is unavailable at online decision time, the policy uses $\widehat{\mathcal{L}}_{\mathrm{det},t}(a)$, an estimate of the action-dependent perception risk. Its construction is introduced in Section~\ref{sec:method}.

The communication overhead is normalized by the high-rate payload $B_H$
\begin{equation}
C_{\mathrm{bit}}(a)
=
\frac{B(a)}{B_H}.
\label{eq:bit_cost}
\end{equation}
Consequently,
$C_{\mathrm{bit}}(a)=0$ for non-transmission actions,
$C_{\mathrm{bit}}(a^{\mathrm{refL}})=B_L/B_H$, and
$C_{\mathrm{bit}}(a^{\mathrm{refH}})=1$. This normalization avoids introducing an additional reference-payload hyperparameter.

Let $R_t$ denote the effective V2X transmission rate at frame $t$ when the link is available. To avoid assigning a transmission term to actions that do not communicate, the action-dependent transmission latency is explicitly written as
\begin{equation}
T_{\mathrm{tx},t}(a)
=
\begin{cases}
0,
&
B(a)=0,
\\[1mm]
\dfrac{B(a)}{R_t},
&
B(a)>0.
\end{cases}
\label{eq:transmission_latency}
\end{equation}
Let $T_{\mathrm{per}}(a)$ denote the perception-processing latency of action $a$. The normalized end-to-end latency cost is then defined as
\begin{equation}
C_{\mathrm{lat},t}(a)
=
\frac{
T_{\mathrm{per}}(a)
+
T_{\mathrm{tx},t}(a)
}{
T_{\max}
},
\label{eq:latency_cost}
\end{equation}
where $T_{\max}$ denotes the maximum allowable end-to-end latency. Using $T_{\max}$ as the normalization factor makes the latency cost directly consistent with the latency constraint introduced below.

Latent imagination introduces additional prediction-dependent risk. The corresponding uncertainty penalty is
\begin{equation}
U_t(a)
=
\begin{cases}
u_t, & a=a^{\mathrm{imag}},\\
0, & \mathrm{otherwise}.
\end{cases}
\label{eq:uncertainty_cost}
\end{equation}
Thus $a^{\mathrm{imag}}$ becomes less preferable when the predicted collaborator representation is unreliable.

To discourage rapid oscillation among operating modes, we additionally define
\begin{equation}
C_{\mathrm{sw}}(a,a_{t-1})
=
\mathbf{1}_{\{a\neq a_{t-1}\}},
\label{eq:switching_cost}
\end{equation}
where $\mathbf{1}_{\{\cdot\}}$ denotes the indicator function and $a_{t-1}$ is the action selected at the preceding frame.

The instantaneous multi-objective cost of candidate action $a$ is
therefore
\begin{equation}
\begin{aligned}
\mathcal{C}_t(a)
=&\;
\widehat{\mathcal{L}}_{\mathrm{det},t}(a)
+
\lambda_b C_{\mathrm{bit}}(a)
+
\lambda_l C_{\mathrm{lat},t}(a)
\\
&+
\lambda_u U_t(a)
+
\lambda_s C_{\mathrm{sw}}(a,a_{t-1}),
\end{aligned}
\label{eq:cost}
\end{equation}
where $\lambda_b$, $\lambda_l$, $\lambda_u$, and $\lambda_s$ are non-negative weighting coefficients controlling communication overhead, latency, prediction uncertainty, and temporal action stability, respectively. The perception-risk term favors actions that preserve detection performance, whereas the remaining terms discourage unnecessary transmission, excessive latency, unreliable imagination,
and unstable switching.

Based on this stage cost, the objective is to learn a
communication--perception policy $\pi_{\phi}(a_t\mid\bm{s}_t)$ parameterized by $\phi$. The resulting
sequential decision problem is formulated as
\begin{equation}
\begin{aligned}
\mathbf{P1}:\quad
\min_{\phi}\quad
&
\mathbb{E}_{\pi_{\phi}}
\left[
\sum_{h=0}^{H_p-1}
\gamma^{h}
\mathcal{C}_{t+h}(a_{t+h})
\right]
\\
\mathrm{s.t.}\quad
\mathrm{C1}:~
&
a_t
\in
\mathcal{A}_{\mathrm{feas}}(\bm{s}_t),
\qquad \forall t,
\\
\mathrm{C2}:~
&
\limsup_{T\rightarrow\infty}
\frac{1}{T}
\sum_{t=1}^{T}
\mathbb{E}
\left[
B(a_t)
\right]
\leq
\bar{B},
\\
\mathrm{C3}:~
&
T_{\mathrm{per}}(a_t)
+
T_{\mathrm{tx},t}(a_t)
\leq
T_{\max},
\qquad \forall t,
\\
\mathrm{C4}:~
&
a_t=a^{\mathrm{imag}}
\Rightarrow
u_t\leq u_{\max},
\qquad \forall t.
\end{aligned}
\label{eq:problemP1}
\end{equation}

In Problem~$\mathbf{P1}$, $H_p$ is the decision horizon,
$\gamma\in(0,1]$ is the temporal discount factor, and $\phi$ denotes the learnable policy parameters. The quantity $\bar{B}$ specifies the allowable long-term average communication payload per decision frame, and $b_t$ in the state represents the instantaneous residual communication budget available at frame $t$. Therefore, $b_t$ and
$\bar{B}$ impose instantaneous and long-term communication-resource requirements, respectively. The parameter $T_{\max}$ specifies the maximum allowable end-to-end latency, while $u_{\max}$ is the maximum predictive uncertainty under which latent imagination is considered admissible.

Constraint~C1 enforces state-dependent causal action feasibility.
Constraint~C2 prevents persistent reliance on communication-intensive
refresh operations. Constraint~C3 imposes the end-to-end latency
requirement, and Constraint~C4 provides a reliability safeguard against
using highly uncertain imagined features. Note that the latency and
uncertainty terms in \eqref{eq:cost} act as \emph{soft preferences}
among admissible actions, whereas C3 and C4 define \emph{hard bounds}
that must not be violated. Therefore, including both the corresponding
cost terms and constraints is not redundant.

Directly solving Problem~$\mathbf{P1}$ remains challenging because the current collaborator observation is unavailable during V2X outages, the action-dependent perception risk is governed by a nonlinear cooperative detector, and successive decisions are temporally coupled through communication conditions, memory, uncertainty, and resource consumption. To address these challenges, Section~\ref{sec:method}
develops \method{}, which combines causal temporal memory and a relative-motion-conditioned latent world model to infer unavailable collaborator information and estimate decision-relevant quantities. An adaptive neuro-fuzzy controller then selects feasible communication--perception actions under the constraints of Problem~$\mathbf{P1}$.

\section{Proposed Agentic LinkDreamer Framework}
\label{sec:method}

\subsection{Overview of Agentic LinkDreamer}
\label{sec:method_overview}

\begin{figure*}[t]
    \centering
    \includegraphics[width=0.98\textwidth]
    {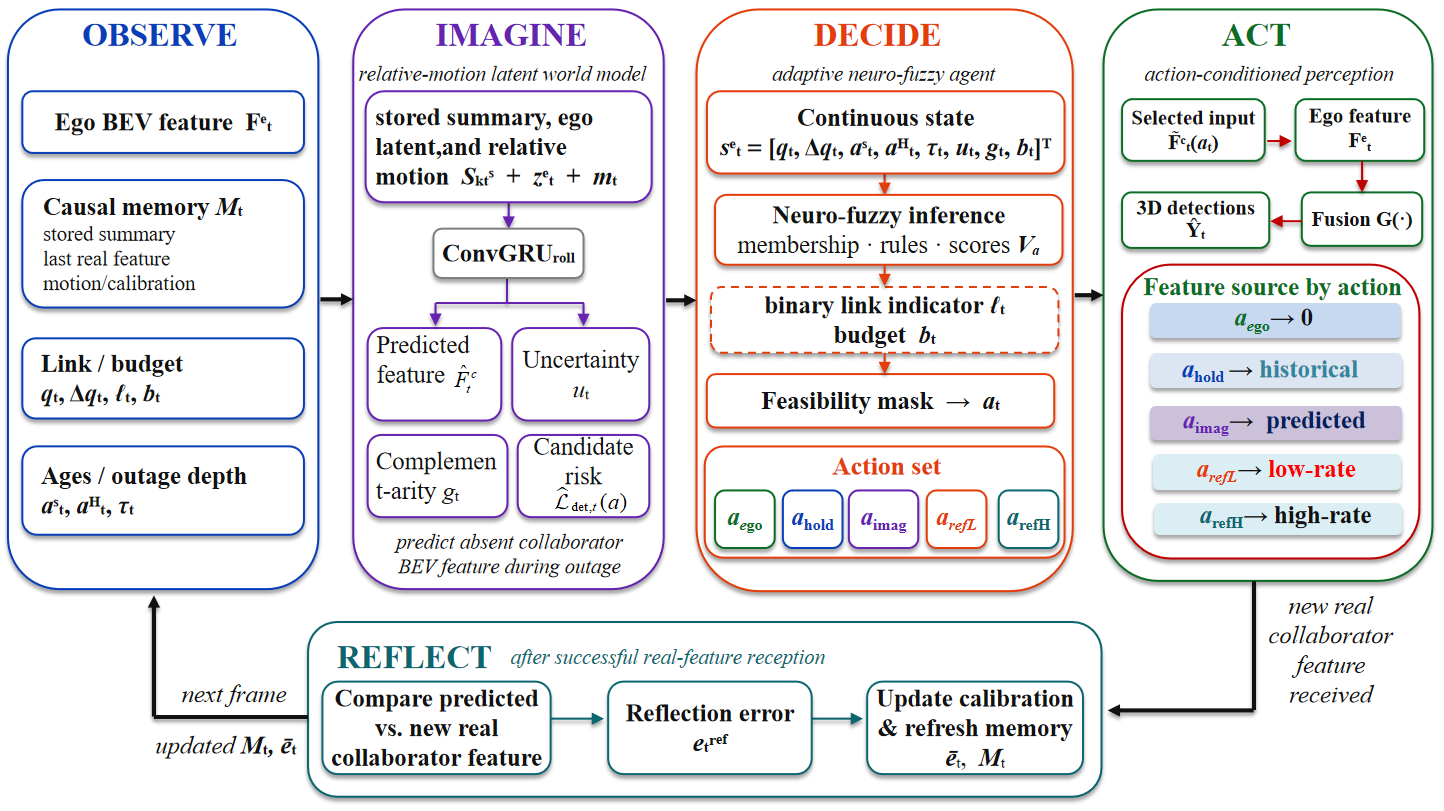}
    \caption{Overall architecture of \method{}.}
    \label{fig:architecture}
\end{figure*}

Fig.~\ref{fig:architecture} illustrates the overall architecture of \method{}. The proposed framework addresses intermittent V2X cooperative perception by integrating causal temporal memory, a relative-motion-conditioned latent world model, uncertainty and perception-utility estimation, and an interpretable neuro-fuzzy decision controller.

Instead of using a fixed compensation strategy to handle missing collaborator information, Agentic LinkDreamer enables the ego vehicle to act as an adaptive perception agent. At each frame, the agent observes the current perception and communication states, predicts the unavailable collaborator
representation when necessary, assesses its reliability and expected perception benefit, and selects an appropriate communication--perception action.

The above procedure is organized as an
\emph{observe--imagine--decide--act--reflect} loop. In the \emph{observe} stage, the ego vehicle forms the decision state from its current BEV feature, causal collaborator memory, relative-motion
information, and V2X link condition. When current collaborator information is unavailable, the \emph{imagine} stage employs the latent world model to infer the missing collaborator representation
and quantify its prediction uncertainty. The \emph{decide} stage then maps the resulting state to an admissible action through the neuro-fuzzy controller. In the \emph{act} stage, the detector input is constructed according to the selected mode, including ego-only perception, historical feature reuse, latent imagination, or communication refresh. Once new collaborator information becomes available, the \emph{reflect} stage compares the predicted and real features to recalibrate uncertainty and update the temporal memory.

\subsection{Causal Temporal Memory and Latent World Model}
\label{sec:world_model}
\label{sec:bev_memory}

\paragraph{Latent BEV encoding}
Based on the BEV features $F_t^i$ defined in
Section~\ref{sec:partial_observability}, a lightweight latent encoder $E_{\ell}(\cdot)$ is introduced for temporal modeling. Specifically, $z_t^i=E_{\ell}(F_t^i)$, where $z_t^i\in\mathbb{R}^{C_{\ell}\times H_{\ell}\times W_{\ell}}$, $C_{\ell}$ is the latent channel dimension and
$H_{\ell}<H$ and $W_{\ell}<W$ are the compressed spatial dimensions. The compact representation $z_t^i$ is used only for temporal-memory construction and world-model rollout. Cooperative feature fusion and object detection remain in the original BEV feature space $\mathbb{R}^{C\times H\times W}$. In our implementation, the BEV features are extracted using a PointPillars backbone~\cite{lang2019pointpillars}.

\paragraph{Causal temporal summary}
The collaborator continuously maintains a compact summary of its local observation history. Given the collaborator latent feature $z_k^c=E_{\ell}(F_k^c)$ at frame $k$, the summary state is recursively updated by
\begin{equation}
h_k^{\mathrm{sum}}
=
\operatorname{ConvGRU}_{\mathrm{sum}}
\left(
z_k^c,
h_{k-1}^{\mathrm{sum}}
\right),
\label{eq:summary_update}
\end{equation}
where $\operatorname{ConvGRU}_{\mathrm{sum}}(\cdot)$ denotes the convolutional recurrent unit used for temporal aggregation and $h_k^{\mathrm{sum}}$ is its hidden state. The transmittable temporal summary is defined as $S_k=h_k^{\mathrm{sum}}$. The update in \eqref{eq:summary_update} depends only on collaborator observations available up to frame $k$ and therefore satisfies causal operation without requiring knowledge of future V2X outages. As defined in Section~\ref{sec:partial_observability}, the ego vehicle only stores the most recently received summary $S_{\kappa_t^S}$ rather than the current local summary $S_t$ when no refresh is available. Its freshness is explicitly represented by the summary age $a_t^S=t-\kappa_t^S$ in \eqref{eq:summary_age}.

The low-rate and high-rate refresh actions defined in
\eqref{eq:actions} determine when the ego-side temporal memory is
updated. Their communication feasibility and payload costs are already
captured by \eqref{eq:feasible_actions} and the multi-objective
decision problem in Section~\ref{sec:problem_formulation}; therefore,
the world model itself does not assume that additional information can
be transmitted after a communication outage has occurred.

\paragraph{Relative-motion-conditioned latent rollout}
When the current collaborator representation is unavailable, directly
reusing the latest received feature ignores scene evolution and
inter-vehicle motion. To predict the unavailable collaborator
representation, \method{} performs a causal latent rollout conditioned
on the latest received temporal summary, the current ego observation,
and relative-motion information.

The motion-conditioning vector is defined as
\begin{equation}
\bm{m}_t
=
\left[
\Delta x_t,\,
\Delta y_t,\,
\Delta\psi_t,\,
\Delta v_t,\,
\tau_t
\right]^{\top},
\label{eq:motion_state}
\end{equation}
where $\Delta x_t$ and $\Delta y_t$ denote the estimated relative
translations between the two agents, $\Delta\psi_t$ is their relative
yaw variation, $\Delta v_t$ denotes the relative velocity variation,
and $\tau_t$ is the current outage depth defined in
\eqref{eq:outage_depth}. These motion variables are estimated causally
from the latest collaborator pose and motion metadata stored in $M_t$
together with the current ego motion; hence, the rollout does not
require the unavailable current collaborator observation.

A lightweight multilayer perceptron $\psi_m(\cdot)$ embeds
$\bm{m}_t$, while another embedding function $\psi_a(\cdot)$ encodes
the temporal-summary age $a_t^S$. At the first imagination step, the
rollout state is initialized as
\begin{equation}
h_t^{\mathrm{init}}
=
P_{\mathrm{init}}
\left(
S_{\kappa_t^S},
\psi_a(a_t^S)
\right),
\label{eq:rollout_init}
\end{equation}
where $P_{\mathrm{init}}(\cdot)$ is a lightweight projection module
that maps the received temporal summary and its age information into
the hidden-state space of the rollout model. Conditioning the
initialization on $a_t^S$ allows the model to distinguish recently
refreshed temporal context from stale temporal memory.

The current ego feature provides additional live scene information.
The rollout state is updated by
\begin{equation}
h_t^{\mathrm{roll}}
=
\operatorname{ConvGRU}_{\mathrm{roll}}
\left(
\left[
E_{\ell}(F_t^e),\,
\psi_m(\bm{m}_t)
\right],
h_{t-1}^{\mathrm{roll}}
\right),
\label{eq:rollout}
\end{equation}
where $\operatorname{ConvGRU}_{\mathrm{roll}}(\cdot)$ denotes the
motion-conditioned recurrent world model. Before concatenation, the
motion embedding $\psi_m(\bm{m}_t)$ is spatially broadcast to match
the dimensions of the ego latent feature. At the first rollout step,
$h_{t-1}^{\mathrm{roll}}$ in \eqref{eq:rollout} is replaced by
$h_t^{\mathrm{init}}$ from \eqref{eq:rollout_init}; during a continuing
outage, the previous rollout state is recursively reused.

A feature decoder $D_f(\cdot)$ maps the latent rollout state back to
the original collaborator BEV feature domain $\widehat{F}_t^c=
D_f\left(h_t^{\mathrm{roll}}\right)$, where $\widehat{F}_t^c\in\mathbb{R}^{C\times H\times W}$, and $\widehat{F}_t^c$ is the predicted collaborator feature before
cross-agent spatial alignment.

\paragraph{Geometry-consistent alignment and historical reuse}
During an outage, the current collaborator pose is unavailable.
Accordingly, let $\widehat{\mathbf{T}}_{c\rightarrow e,t}$ denote the
collaborator-to-ego transformation estimated from the latest available collaborator pose and motion information in $M_t$ together with the current ego pose. The predicted collaborator feature is transformed into the current ego coordinate frame as
\begin{equation}
\widehat{\bar{F}}_t^c
=
\mathcal{T}
\left(
\widehat{F}_t^c;
\widehat{\mathbf{T}}_{c\rightarrow e,t}
\right),
\label{eq:pred_aligned_feature}
\end{equation}
where $\mathcal{T}(\cdot)$ is the same spatial transformation operator
defined in \eqref{eq:aligned_feature}.

For the historical-holding action, the latest high-fidelity
collaborator feature $F_{\kappa_t}^{c}$ may have been received several frames earlier. Its staleness is quantified by
$a_t^H=t-\kappa_t$ in \eqref{eq:feature_age}. Before fusion, the stale feature is first compensated for temporal motion and then aligned with the current ego frame
\begin{equation}
\bar{F}_t^{c,\mathrm{hold}}
=
\mathcal{T}
\left(
\mathcal{W}_{\kappa_t\rightarrow t}
\left(
F_{\kappa_t}^{c}
\right);
\widehat{\mathbf{T}}_{c\rightarrow e,t}
\right),
\label{eq:hold_aligned_feature}
\end{equation}
where $\mathcal{W}_{\kappa_t\rightarrow t}(\cdot)$ denotes temporal
motion compensation over the $a_t^H$-frame interval between the last
high-fidelity reception and the current frame. Thus, $\mathcal{W}_{\kappa_t\rightarrow t}(\cdot)$ compensates the temporal
staleness of the stored feature, whereas $\mathcal{T}(\cdot)$ resolves the spatial coordinate difference between the collaborator and ego vehicles.

The two resulting representations, $\widehat{\bar{F}}_t^c$ and
$\bar{F}_t^{c,\mathrm{hold}}$, respectively support the $a^{\mathrm{imag}}$ and $a^{\mathrm{hold}}$ actions. Their reliability is further characterized by the uncertainty, complementarity, and action-dependent risk estimators introduced in the following subsection.

\subsection{Decision-Relevant Uncertainty, Complementarity, and Risk Estimation}
\label{sec:uncertainty}

The decision state in \eqref{eq:state} requires the predictive
uncertainty $u_t$ and perceptual complementarity $g_t$, while the
multi-objective cost in \eqref{eq:cost} further relies on the
action-dependent perception risk $\widehat{\mathcal{L}}_{\mathrm{det},t}(a)$.
We therefore construct three lightweight estimators on top of the
latent world-model state to quantify prediction reliability,
collaborator usefulness, and the expected perception consequence of
each candidate action.

\paragraph{Prediction uncertainty}
The world model first produces a spatial predictive variance map
\begin{equation}
\widehat{V}_{t,\mathrm{pix}}
=
\operatorname{softplus}
\left(
D_u(h_t^{\mathrm{roll}})
\right),
\label{eq:pixel_uncertainty}
\end{equation}
where $D_u(\cdot)$ denotes the uncertainty decoder and
$\widehat{V}_{t,\mathrm{pix}}\geq0$.
The mean predictive variance is summarized as
\begin{equation}
v_t^{\mathrm{pred}}
=
\operatorname{mean}
\left(
\widehat{V}_{t,\mathrm{pix}}
\right).
\label{eq:predictive_variance}
\end{equation}

In order to additionally capture epistemic uncertainty, $M_u$ stochastic world-model predictions are generated with dropout enabled. Let
$\widehat{F}_{t,m}^{c}$ denote the $m$-th stochastic prediction, $m\in\{1,\ldots,M_u\}$. The feature-level disagreement is summarized
into a scalar as
\begin{equation}
v_t^{\mathrm{epi}}
=
\operatorname{mean}
\left[
\operatorname{Var}_{m}
\left(
\widehat{F}_{t,m}^{c}
\right)
\right],
\label{eq:epistemic_uncertainty}
\end{equation}
where the outer mean is taken over all feature channels and spatial locations.

The global predictive uncertainty used by the decision controller is also defined as
\begin{equation}
u_t
=
\operatorname{sigmoid}
\left(
\alpha_0
+
\alpha_1 v_t^{\mathrm{pred}}
+
\alpha_2 v_t^{\mathrm{epi}}
+
\alpha_3
\frac{\tau_t}{H_{\max}}
+
\alpha_4
\bar e_{\kappa_t}
\right),
\label{eq:global_uncertainty}
\end{equation}
where $H_{\max}$ denotes the maximum modeled outage horizon, $\bar e_{\kappa_t}$ is the latest available reflection-based calibration statistic, and $\alpha_0,\ldots,\alpha_4$ are learnable calibration parameters. The scalar uncertainty cues are normalized before fusion. The sigmoid mapping guarantees $u_t\in[0,1]$, consistent with the definition in \eqref{eq:state}.

\paragraph{Perceptual complementarity}
Prediction reliability alone does not determine whether remote information should be used. A reliable collaborator representation may
provide little additional benefit when the ego vehicle already observes the same objects or spatial regions. We therefore construct a compact
complementarity descriptor $\bm{\xi}_t^g$ from geometric field-of-view overlap, relative visibility, recent remote-only detection evidence
from valid cooperative frames, and the currently available collaborator memory. The perceptual complementarity is estimated as
\begin{equation}
g_t
=
\operatorname{sigmoid}
\left(
D_g(\bm{\xi}_t^g)
\right),
\label{eq:complementarity}
\end{equation}
where $D_g(\cdot)$ denotes a lightweight complementarity predictor and
$g_t\in[0,1]$.

During offline training, the supervision target is constructed from
the relative perception improvement,
\begin{equation}
g_t^{\star}
=
\operatorname{clip}
\left(
\frac{
\mathcal{L}_{\mathrm{det},t}^{\mathrm{ego}}
-
\mathcal{L}_{\mathrm{det},t}^{\mathrm{coop}}
}{
\mathcal{L}_{\mathrm{det},t}^{\mathrm{ego}}
+\epsilon
},
0,1
\right),
\label{eq:complementarity_target}
\end{equation}
where
$\mathcal{L}_{\mathrm{det},t}^{\mathrm{ego}}$ and
$\mathcal{L}_{\mathrm{det},t}^{\mathrm{coop}}$ denote the detection
losses under ego-only and valid cooperative perception, respectively,
and $\epsilon>0$ is a small numerical constant.
A large $g_t$ indicates that collaborator information is expected to
provide substantial additional perception benefit, whereas a small
value indicates strong redundancy with the ego observation.

\paragraph{Action-dependent perception risk}
While $u_t$ and $g_t$ characterize prediction reliability and
collaborator usefulness, respectively, the controller must also
estimate the downstream perception consequence of each candidate
action. A compact world-model descriptor is first obtained as
\begin{equation}
r_t
=
\operatorname{GAP}
\left(
h_t^{\mathrm{roll}}
\right),
\label{eq:risk_descriptor}
\end{equation}
where $\operatorname{GAP}(\cdot)$ denotes global average pooling.

For each candidate action
$a\in\mathcal{A}_{\mathrm{feas}}(\bm{s}_t)$, the predicted detection
risk is
\begin{equation}
\widehat{\mathcal{L}}_{\mathrm{det},t}(a)
=
\operatorname{softplus}
\left(
D_r
\left(
[
r_t,\,
\bm{s}_t,\,
\bm{e}_a
]
\right)
\right),
\label{eq:predicted_detection_risk}
\end{equation}
where $D_r(\cdot)$ denotes the risk-estimation network and
$\bm{e}_a\in\{0,1\}^{|\mathcal{A}|}$ is the one-hot embedding of
action $a$. The softplus output ensures a non-negative predicted
detection risk.

Consequently, $u_t$ measures whether the predicted collaborator
representation is reliable, $g_t$ measures whether collaborator
information is expected to be useful, and
$\widehat{\mathcal{L}}_{\mathrm{det},t}(a)$ estimates the perception
consequence of executing each candidate action. These quantities form
the decision-relevant interface between the latent world model and the
neuro-fuzzy controller introduced in the following subsection.

\subsection{Adaptive Interpretable Neuro-Fuzzy Decision Controller}
\label{sec:fuzzy}

The latent world model provides the predictive information required for communication--perception decision making. The uncertainty $u_t$ and perceptual complementarity $g_t$ estimated in the previous subsection are incorporated into the decision state $\bm{s}_t$, while the action-dependent perception risk $\widehat{\mathcal{L}}_{\mathrm{det},t}(a)$ provides cost supervision during policy optimization. To approximate the constrained policy in Problem~$\mathbf{P1}$ while retaining interpretability, we employ a differentiable Takagi--Sugeno neuro-fuzzy controller.

 The neuro-fuzzy controller takes the normalized continuous decision
descriptor $\widetilde{\bm{s}}_t$ defined in
\eqref{eq:continuous_state} as input, while the binary link indicator $\ell_t$ is handled separately through action-feasibility masking. Accordingly, the controller operates on $d=8$ continuous decision variables.
\begin{equation}
\mu_{j,m}(\widetilde{s}_{t,j})
=
\exp\left[
-\frac{
\left(
\widetilde{s}_{t,j}-c_{j,m}
\right)^2
}{
2\sigma_{j,m}^2
}
\right],
\label{eq:membership}
\end{equation}
where $\widetilde{s}_{t,j}$ denotes the $j$-th normalized continuous decision variable, $c_{j,m}$ and $\sigma_{j,m}>0$ are the learnable center and width of the $m$-th Gaussian fuzzy set for input dimension $j$.

For fuzzy rule $r$, the firing strength is computed as
\begin{equation}
\omega_r(\widetilde{\bm{s}}_t)
=
\prod_{j\in\mathcal{J}_r}
\mu_{j,m_{r,j}}
\left(
\widetilde{s}_{t,j}
\right),
\label{eq:rule_strength}
\end{equation}
where $\mathcal{J}_r$ denotes the subset of continuous state dimensions involved in rule $r$, and $m_{r,j}$ specifies the linguistic fuzzy set assigned to dimension $j$ by that rule. To enable data-driven rule pruning, each rule is further associated with a non-negative learnable gate $\xi_r=\operatorname{softplus}(\zeta_r)$, where $\zeta_r$ is an unconstrained learnable parameter and $\xi_r\geq0$ represents the effective importance of rule $r$. The gated firing strengths are normalized as
\begin{equation}
\bar{\omega}_r(\widetilde{\bm{s}}_t)
=
\frac{
\xi_r
\omega_r(\widetilde{\bm{s}}_t)
}{
\sum_{r'}
\xi_{r'}
\omega_{r'}(\widetilde{\bm{s}}_t)
+\epsilon
},
\label{eq:normalized_rule_strength}
\end{equation}
where $\epsilon>0$ is a small constant introduced for numerical stability.

For each candidate action $a$, rule $r$ employs a first-order Takagi--Sugeno consequent
\begin{equation}
y_{r,a}(\widetilde{\bm{s}}_t)
=
\beta_{r,a,0}
+
\bm{\beta}_{r,a}^{\top}
\widetilde{\bm{s}}_t,
\label{eq:ts_consequent}
\end{equation}
where $\beta_{r,a,0}$ is the bias term and
$\bm{\beta}_{r,a}\in\mathbb{R}^{d}$ contains the first-order Takagi--Sugeno consequent coefficients for rule $r$ and action $a$. The resulting action score is
\begin{equation}
V_a(\widetilde{\bm{s}}_t)
=
\sum_r
\bar{\omega}_r(\widetilde{\bm{s}}_t)
y_{r,a}(\widetilde{\bm{s}}_t),
\label{eq:action_score}
\end{equation}

To enforce all instantaneous constraints in Problem~$\mathbf{P1}$,
the link- and budget-feasible set
$\mathcal{A}_{\mathrm{feas}}(\bm{s}_t)$ defined in
\eqref{eq:feasible_actions} is further filtered according to the
latency and imagination-reliability constraints. The resulting
admissible action set is
\begin{equation}
\begin{aligned}
\mathcal{A}_{\mathrm{adm}}(\bm{s}_t)
=
\Big\{
a\in\mathcal{A}_{\mathrm{feas}}(\bm{s}_t):
&
~T_{\mathrm{per}}(a)
+
T_{\mathrm{tx},t}(a)
\leq T_{\max},
\\
&
~a\neq a^{\mathrm{imag}}
~\text{or}~
u_t\leq u_{\max}
\Big\}.
\end{aligned}
\label{eq:admissible_actions}
\end{equation}
Thus, $\mathcal{A}_{\mathrm{feas}}(\bm{s}_t)$ enforces instantaneous
V2X-link and communication-budget feasibility, whereas
$\mathcal{A}_{\mathrm{adm}}(\bm{s}_t)$ additionally enforces the
latency constraint C3 and the imagination-reliability constraint C4.

The action scores are converted into a differentiable policy over the admissible actions
\begin{equation}
\pi_{\phi}(a\mid\bm{s}_t)
=
\frac{
\exp
\left(
V_a(\widetilde{\bm{s}}_t)/\eta_{\pi}
\right)
}{
\displaystyle
\sum_{a'\in\mathcal{A}_{\mathrm{adm}}(\bm{s}_t)}
\exp
\left(
V_{a'}(\widetilde{\bm{s}}_t)/\eta_{\pi}
\right)
},
\quad
a\in\mathcal{A}_{\mathrm{adm}}(\bm{s}_t),
\label{eq:fuzzy_policy}
\end{equation}
where $\eta_{\pi}>0$ is the policy temperature and $\phi$ collects
the learnable membership-function, rule-gating, and Takagi--Sugeno
consequent parameters. Actions outside
$\mathcal{A}_{\mathrm{adm}}(\bm{s}_t)$ are assigned zero probability.

During inference, the final action is selected deterministically as
\begin{equation}
a_t=\arg\max_{
a\in\mathcal{A}_{\mathrm{adm}}(\bm{s}_t)
}
V_a(\widetilde{\bm{s}}_t).
\label{eq:action_selection}
\end{equation}

The initial fuzzy-rule base provides interpretable decision priors
rather than hard-coded action assignments. For example, increasing the temporal-summary age $a_t^S$ or predictive uncertainty $u_t$
reduces the preference for latent imagination because both indicate reduced confidence in world-model rollout. In contrast, increasing
the high-fidelity feature age $a_t^H$ reduces the preference for the historical-holding action because the stored collaborator feature
becomes progressively stale. A large perceptual complementarity $g_t$ increases the value of collaborator-assisted actions, whereas
low complementarity suppresses unnecessary communication. When the V2X link is available, stale memory together with sufficient
communication budget promotes low- or high-rate refresh depending on the expected perception benefit and communication quality.

\subsection{Action-Conditioned Perception and Reflection}
\label{sec:action_reflection}

\paragraph{Action-specific feature construction}
After an action $a_t$ is selected, the corresponding collaborator representation is constructed and supplied to the cooperative detector. For the two communication-refresh actions, let $\mathcal{Q}_L(\cdot)$ and $\mathcal{Q}_H(\cdot)$ denote the low- and
high-rate communication encoders, respectively. Their reconstructed collaborator representations are
\begin{align}
F_t^{c,L}
&=
D_L
\left(
\mathcal{Q}_L(S_t)
\right),
\label{eq:lowrate_feature}
\\
F_t^{c,H}
&=
D_H
\left(
\mathcal{Q}_H(F_t^c)
\right),
\label{eq:highrate_feature}
\end{align}
where $D_L(\cdot)$ and $D_H(\cdot)$ denote the corresponding decoders. The low-rate branch reconstructs a collaborator representation from
the compressed temporal summary, whereas the high-rate branch provides a higher-fidelity representation of the current collaborator feature,
with $B_H>B_L$.

Since both refresh actions are executed only when the V2X link is available, their reconstructed features are aligned to the current ego
coordinate frame using
\begin{equation}
\bar F_t^{c,r}
=
\mathcal{T}
\left(
F_t^{c,r};
\mathbf{T}_{c\rightarrow e,t}
\right),
\label{eq:refresh_aligned}
\end{equation}
where $r\in\{L,H\}$ denotes the low- or high-rate refresh branch. For latent imagination, a spatial confidence map is constructed from
the predicted variance
\begin{equation}
c_t
=
\exp
\left(
-\widehat{V}_{t,\mathrm{pix}}/\beta_u
\right),
\label{eq:spatial_confidence}
\end{equation}
where $\beta_u>0$ controls the uncertainty attenuation. Because the current collaborator pose is unavailable during an outage, the confidence map is aligned using the estimated relative transformation
\begin{equation}
\bar c_t
=
\mathcal{T}
\left(
c_t;
\widehat{\mathbf{T}}_{c\rightarrow e,t}
\right).
\label{eq:aligned_confidence}
\end{equation}

The action-conditioned collaborator input is therefore
\begin{equation}
\widetilde{F}_t^c(a_t)
=
\begin{cases}
\mathbf{0},
& a_t=a^{\mathrm{ego}},
\\[1.5mm]
\bar F_t^{c,\mathrm{hold}},
& a_t=a^{\mathrm{hold}},
\\[1.5mm]
\bar c_t\odot\widehat{\bar F}_t^c,
& a_t=a^{\mathrm{imag}},
\\[1.5mm]
\bar F_t^{c,L},
& a_t=a^{\mathrm{refL}},
\\[1.5mm]
\bar F_t^{c,H},
& a_t=a^{\mathrm{refH}},
\end{cases}
\label{eq:featureaction}
\end{equation}
where $\odot$ denotes element-wise multiplication. The resulting detection output is obtained using the same cooperative fusion module and detection head defined in
\eqref{eq:cooperative_detection},
\begin{equation}
\widehat{Y}_t
=
G
\left(
F_t^e,
\widetilde{F}_t^c(a_t)
\right).
\label{eq:action_detection}
\end{equation}
Thus, all operating modes share the same downstream detector and differ only in the collaborator representation supplied to the fusion module.

\paragraph{Reflection after link recovery}
When a new high-fidelity collaborator feature is successfully received after an outage, the ego agent compares the previously predicted representation with the corresponding real feature. The discrepancy in the aligned BEV feature domain is measured as
\begin{equation}
e_t^{\mathrm{ref}}
=
\frac{1}{CHW}
\left\|
\widehat{\bar F}_t^c
-
\bar F_t^c
\right\|_1,
\label{eq:reflection_error}
\end{equation}
where $\bar F_t^c$ denotes the real collaborator feature aligned to the current ego coordinate frame.

An exponentially smoothed calibration statistic is maintained as
\begin{equation}
\bar e_t
=
\rho\bar e_{t-1}
+
(1-\rho)e_t^{\mathrm{ref}},
\label{eq:reflection_memory}
\end{equation}
where $0\leq\rho<1$, which controls the update rate of the calibration memory. The updated statistic $\bar e_t$ is subsequently incorporated into the
uncertainty estimate in \eqref{eq:global_uncertainty}. The newly received high-fidelity feature and its associated temporal summary also replace the corresponding stale memory entries. In this way, reflection updates the temporal memory and uncertainty calibration without modifying the BEV perception backbone online.

\subsection{Inference, Training, and Computational Complexity}
\label{sec:training}
\paragraph{Causal inference}
Algorithm~\ref{alg:inference} summarizes the online decision procedure of \method{}. At each frame, the agent uses only the current observation and causal collaborator memory to estimate the decision state, evaluate admissible
actions, and construct the collaborator representation for cooperative perception.

\begin{algorithm}[t]
\caption{Causal inference with \method{}}
\label{alg:inference}
\begin{algorithmic}[1]
\Require $X_t^e$, link state $\ell_t$, memory $M_{t-1}$, previous action $a_{t-1}$

\State Encode ego feature $F_t^e \gets E_{\mathrm{bev}}(X_t^e)$
\State Update $q_t,\Delta q_t,a_t^S,a_t^H,\tau_t,b_t$ and relative motion $\bm{m}_t$
\State Roll the world model to obtain $\widehat{F}_t^c$ and $h_t^{\mathrm{roll}}$
\State Estimate uncertainty $u_t$, complementarity $g_t$, and
$\widehat{L}_{\mathrm{det},t}(a)$
\State Form $\widetilde{\bm{s}}_t$ and construct
$\mathcal{A}_{\mathrm{adm}}(\bm{s}_t)$
\State Compute fuzzy utilities $V_a(\widetilde{\bm{s}}_t)$
\State $a_t \gets
\arg\max_{a\in\mathcal{A}_{\mathrm{adm}}(\bm{s}_t)}
V_a(\widetilde{\bm{s}}_t)$

\State Construct action-conditioned collaborator feature
$\widetilde{F}_t^c(a_t)$ according to Eq.~(\ref{eq:featureaction})
\State $\widehat{Y}_t \gets
G(F_t^e,\widetilde{F}_t^c(a_t))$

\If{a communication refresh is executed}
    \State Update the corresponding entries of causal memory $M_t$
\EndIf
\If{a new high-fidelity feature is received after an outage}
    \State Update $e_t^{\mathrm{ref}}$, $\bar e_t$, and $M_t$
\EndIf

\State \Return $\widehat{Y}_t$, $a_t$, $u_t$, $M_t$
\end{algorithmic}
\end{algorithm}
\paragraph{Progressive training}
The complete framework is trained progressively to stabilize the interaction among cooperative perception, latent prediction, decision-relevant estimation, and neuro-fuzzy policy learning. In Stage~I, the cooperative perception backbone, fusion module, and detection head are pretrained on link-available samples using
\begin{equation}
\mathcal{L}_{\mathrm{det}}
=
\mathcal{L}_{\mathrm{cls}}
+
\lambda_{\mathrm{reg}}
\mathcal{L}_{\mathrm{reg}},
\label{eq:det_loss}
\end{equation}
where $\mathcal{L}_{\mathrm{cls}}$ and
$\mathcal{L}_{\mathrm{reg}}$ denote the classification and bounding-box regression losses, respectively, and $\lambda_{\mathrm{reg}}$ balances the two terms. After convergence, the BEV backbone is temporarily frozen. In Stage~II, the latent world model and uncertainty estimator
are trained using temporal sequences containing valid collaborator context followed by a simulated outage of length
$1\leq H\leq H_{\max}$. Their objective is
\begin{equation}
\begin{aligned}
\mathcal{L}_{\mathrm{wm}}
=
\frac{1}{H}
\sum_{h=1}^{H}
\Bigg[
\frac{1}{CHW}
\sum_{c,p}
\Bigg(
&
\frac{
\left(
\widehat{F}_{t+h,c,p}^{c}
-
F_{t+h,c,p}^{c}
\right)^2
}{
2\left(
\widehat{V}_{t+h,p}
+\epsilon
\right)
}
\\
&+
\frac{1}{2}
\log
\left(
\widehat{V}_{t+h,p}
+\epsilon
\right)
\Bigg)
+
\lambda_{\mathrm{dyn}}
\mathcal{L}_{\mathrm{dyn},t+h}
\Bigg],
\end{aligned}
\label{eq:wm_loss}
\end{equation}
where $c$ and $p$ index the feature channel and spatial BEV position, respectively, $\widehat{V}_{t+h,p}$ is the predicted variance, $\epsilon>0$ is a numerical-stability constant, $\mathcal{L}_{\mathrm{dyn},t+h}$ regularizes the
relative-motion-conditioned latent dynamics, and
$\lambda_{\mathrm{dyn}}$ controls its contribution. The outage length and temporal-summary age are randomized during training to improve robustness to different rollout depths and memory staleness.

In Stage~III, the fusion module and detection head are fine-tuned using a balanced mixture of the communication--perception modes defined in Section~\ref{sec:action_reflection}, while the BEV backbone and latent dynamics remain fixed. Since the ground-truth annotations and collaborator observations are available offline, the candidate actions in $\mathcal{A}_{\mathrm{feas}}(\bm{s}_t)$ can be evaluated on the same scene. This provides supervision for the action-risk and complementarity
estimators:
\begin{equation}
\mathcal{L}_{\mathrm{risk}}
=
\frac{
1
}{
|\mathcal{A}_{\mathrm{feas}}(\bm{s}_t)|
}
\sum_{a\in\mathcal{A}_{\mathrm{feas}}(\bm{s}_t)}
\left|
\widehat{\mathcal{L}}_{\mathrm{det},t}(a)
-
\mathcal{L}_{\mathrm{det},t}(a)
\right|,
\label{eq:risk_loss}
\end{equation}
and
\begin{equation}
\mathcal{L}_{\mathrm{comp}}
=
\left(
g_t-g_t^{\star}
\right)^2,
\label{eq:complementarity_loss}
\end{equation}
where $\mathcal{L}_{\mathrm{det},t}(a)$ denotes the offline detection loss associated with action $a$, and $g_t^{\star}$ is the complementarity target defined in \eqref{eq:complementarity_target}.

In Stage~IV, the adaptive neuro-fuzzy controller is optimized according to Problem~$\mathbf{P1}$. For each admissible candidate action, the finite-horizon cost-to-go is estimated as
\begin{equation}
\widehat{Q}_t(a)
=
\mathbb{E}
\left[
\left.
\sum_{h=0}^{H_p-1}
\gamma^h
\mathcal{C}_{t+h}(a_{t+h})
\right|
\bm{s}_t,\,
a_t=a
\right],
\label{eq:cost_to_go}
\end{equation}
where the decision state and causal memory evolve according to the selected actions and the recorded link-availability sequence, while unavailable collaborator observations are supplied by the latent world model. The policy is optimized only over the admissible action set
$\mathcal{A}_{\mathrm{adm}}(\bm{s}_t)$ defined in
\eqref{eq:admissible_actions}:
\begin{equation}
\begin{aligned}
\mathcal{L}_{\mathrm{agent}}
=
&
\sum_{a\in\mathcal{A}_{\mathrm{adm}}(\bm{s}_t)}
\pi_{\phi}(a\mid\bm{s}_t)
\widehat{Q}_t(a)
-
\lambda_H
\mathcal{H}(\pi_{\phi})
\\
&
+
\lambda_R
\|\bm{\xi}\|_1
+
\lambda_M
\mathcal{L}_{\mathrm{mono}}
+
\nu_B
\left(
\widehat{B}_{\phi}
-
\bar B
\right),
\end{aligned}
\label{eq:agent_loss}
\end{equation}
where $\mathcal{H}(\pi_{\phi})$ denotes the policy entropy,
$\bm{\xi}$ collects the learnable fuzzy-rule gates,
$\mathcal{L}_{\mathrm{mono}}$ is the monotonicity regularizer, and $\lambda_H$, $\lambda_R$, and $\lambda_M$ control exploration, rule sparsity, and monotonicity, respectively. The dual variable $\nu_B\geq0$ enforces the long-term communication constraint through
the expected payload
\begin{equation}
\widehat{B}_{\phi}
=
\mathbb{E}_{\bm{s}_t}
\left[
\sum_{a\in\mathcal{A}_{\mathrm{adm}}(\bm{s}_t)}
\pi_{\phi}(a\mid\bm{s}_t)
B(a)
\right],
\label{eq:expected_payload}
\end{equation}
and is updated by projected gradient ascent $\nu_B
\leftarrow
\left[
\nu_B
+
\eta_B
\left(
\widehat{B}_{\phi}
-
\bar B
\right)
\right]_{+}$
where $\eta_B>0$ is the dual-update step size and
$[x]_{+}=\max(x,0)$. The instantaneous link, communication-budget,
latency, and imagination-reliability constraints are therefore enforced
through $\mathcal{A}_{\mathrm{adm}}(\bm{s}_t)$, whereas the long-term
average communication constraint is handled by the dual update.

After the four progressive stages, all trainable modules are jointly
refined using a reduced learning rate:
\begin{equation}
\begin{aligned}
\mathcal{L}_{\mathrm{joint}}
=
&
\mathcal{L}_{\mathrm{det}}
+
\lambda_{\mathrm{wm}}
\mathcal{L}_{\mathrm{wm}}
+
\lambda_{\mathrm{risk}}
\mathcal{L}_{\mathrm{risk}}
\\
&
+
\lambda_{\mathrm{comp}}
\mathcal{L}_{\mathrm{comp}}
+
\lambda_{\mathrm{agent}}
\mathcal{L}_{\mathrm{agent}},
\end{aligned}
\label{eq:joint_loss}
\end{equation}
where $\lambda_{\mathrm{wm}}$, $\lambda_{\mathrm{risk}}$,
$\lambda_{\mathrm{comp}}$, and $\lambda_{\mathrm{agent}}$ balance the
world-model, risk-estimation, complementarity, and policy objectives relative to the detection objective. Gradients induced solely by communication-cost terms are stopped before the BEV backbone so that semantic perception features are not modified merely to reduce communication overhead. Quantization-aware training is applied only to the representations transmitted through the communication branches.



\section{Experiments}
\label{sec:experiments}

This section evaluates Agentic LinkDreamer under intermittent V2X communication using the real-world V2X-Real benchmark. The experimental
evaluation focuses on four aspects: cooperative-perception recovery when the current collaborator observation becomes unavailable, robustness of latent imagination to increasing outage duration, the communication--perception
trade-off achieved by adaptive decision making, and the individual contributions of the world-model and neuro-fuzzy decision components.

\subsection{Experimental Setup}
\label{sec:exp_setup}

\subsubsection{Dataset and Temporal Sequence Construction}

Experiments are conducted on V2X-Real, a real-world multi-agent cooperative perception dataset containing synchronized LiDAR observations from connected vehicles and intelligent infrastructure. We use the
vehicle-to-vehicle (V2V) split as the main benchmark because both the ego vehicle and the collaborator are moving agents. This setting is particularly suitable for evaluating the proposed relative-motion-conditioned world model, since missing collaborator information must be reconstructed under both temporal scene evolution and inter-vehicle motion.

World-model learning requires temporally consistent collaborator identities.
The original frame-level samples are therefore reorganized into fixed-agent
temporal sequences $(s,e,c,t-K+1:t+H)$, where $s$ denotes the scenario, $e$ and $c$ denote the ego and collaborator identities, respectively, $K$ is the historical observation length, and $H$ denotes the prediction horizon. A valid sequence does not cross scenario
boundaries and maintains the same ego--collaborator pair throughout the
temporal window.

Training, validation, and test sets are separated at the scenario level to avoid information leakage between temporally adjacent frames. Unless otherwise specified, $K=4$ historical collaborator frames are used to construct the causal temporal summary, while the prediction horizon is sampled from $H \in \{1,2,4,6,8,10\}$.

\subsubsection{Intermittent V2X Communication Protocol}

V2X-Real provides synchronized multi-agent perception measurements but does not contain physical-layer channel measurements or frame-level packet-delivery traces. We therefore employ the geometry-derived communication degradation protocol introduced in Section~3 to generate reproducible V2X outage episodes.

Let $d_t$ denote the planar distance between the ego vehicle and the collaborator. Geometric blockage is represented by $o_t$, defined as the
normalized density of non-ground LiDAR returns within a corridor surrounding the line segment between the two sensing agents. Endpoint regions are excluded to avoid counting structures belonging to the sensing platforms themselves. The resulting geometry-derived link-quality proxy is
\begin{equation}
q_t =
\sigma\left(
\frac{d_0-d_t}{s_d}
\right)
\exp\left(
-\frac{o_t}{o_0}
\right),
\end{equation}
where $d_0$, $s_d$, and $o_0$ determine the sensitivity to distance and geometric blockage.

A hysteresis mechanism converts $q_t$ into the binary link state $\ell_t$. The link becomes unavailable when the proxy remains below a predefined lower threshold for several consecutive frames and becomes available again only after exceeding a higher recovery threshold. All protocol parameters are
selected using the validation scenarios and remain fixed during testing. The protocol is strictly causal. At frame $t$, the agent has access only to the current and historical communication states, while future link states are never exposed to the decision controller. Communication-refresh actions are therefore executable only when the current link state and communication budget satisfy the feasibility conditions defined in Section~3.

\subsubsection{Compared Methods}

All compared methods use the same PointPillars-based BEV encoder,
intermediate-fusion module, and detection head unless otherwise specified.
They differ in the strategy used to handle missing collaborator information
or allocate communication resources.

\paragraph{Cooperative upper reference}
The current collaborator feature is assumed to be successfully received at
every frame. This setting represents the empirical upper reference for
cooperative perception.

\paragraph{Ego-only}
The detector operates without remote collaborator information and uses only
the ego-vehicle BEV representation.

\paragraph{Degraded cooperation}
Normal cooperative fusion is used while the V2X link is available, whereas the collaborator input is removed when an outage occurs. This setting quantifies the degradation introduced when the cooperative detector loses access to the current collaborator observation.
\paragraph{Relative-motion warp}
The most recently received collaborator feature is reused after temporal motion compensation and coordinate alignment. This method represents a non-generative historical-feature recovery strategy.
\paragraph{Always-imagine}
The latent world model predicts the missing collaborator feature whenever the link is unavailable, and the imagined representation is directly injected into the cooperative detector without adaptive action selection. This baseline isolates unconditional world-model-based recovery.

\paragraph{Fixed refresh}
Collaborator information is refreshed according to a predefined periodic schedule. Multiple refresh intervals are evaluated to provide operating points with communication budgets comparable to those of Agentic LinkDreamer.

\paragraph{Agentic LinkDreamer}
The proposed method combines causal temporal memory, relative-motion-conditioned latent imagination, uncertainty estimation, perceptual complementarity, and adaptive neuro-fuzzy decision making. At each frame, the controller selects among ego-only perception, historical feature reuse, latent imagination, low-rate refresh, and high-rate refresh according to the
current decision state and admissible-action constraints.

\subsubsection{Evaluation Metrics and Statistical Protocol}

Perception performance is primarily evaluated on communication-outage frames, since these frames directly measure the ability of each recovery method to compensate for missing collaborator information. We report detection AP at IoU thresholds of $0.5$ and $0.7$, denoted by AP@0.5 and AP@0.7 respectively.

To quantify the fraction of the cooperative-perception degradation recovered by a method, we define
\begin{equation}
\mathrm{Recovery}_{\delta}
=
100
\frac{
\mathrm{AP}_{\mathrm{method},\delta}
-
\mathrm{AP}_{\mathrm{degraded},\delta}
}{
\mathrm{AP}_{\mathrm{coop},\delta}
-
\mathrm{AP}_{\mathrm{degraded},\delta}
},
\label{eq:recovery_metric}
\end{equation}
where $\delta \in \{0.5,0.7\}$ denotes the IoU threshold. The degraded baseline corresponds to $0\%$ recovery, whereas the cooperative upper reference corresponds to $100\%$. Recovery is reported only for methods designed to reconstruct or reacquire collaborator information; the ego-only baseline is therefore excluded from this metric.

Communication overhead is measured using the actual action-dependent payload rather than the nominal refresh frequency. The average transmitted data volume is reported in MB/frame. For the adaptive policy, the frequencies of low-rate and high-rate refresh actions can additionally be obtained from thepolicy trajectory.

Average end-to-end latency is used as a system-level feasibility check for communication--perception actions. The experimental evaluation focuses on perception recovery and communication efficiency rather than neural-network acceleration.

For the final comparison and ablation experiments, stochastic training results
are reported over multiple independent runs using the same scenario-level train/validation/test split. Statistical comparisons are performed at the scenario level rather than the individual-frame level to avoid treating temporally correlated frames as independent samples.

\subsection{Overall Performance under V2X Outages}
\label{sec:overall_results}

Table~\ref{tab:main_results} summarizes the outage-frame perception performance on the V2X-Real V2V test split. The cooperative upper reference and degraded-cooperation baseline define the performance gap caused by intermittent collaborator availability, while the recovery methods quantify
how much of this gap can be restored without continuous access to the current collaborator observation. For the communication-aware methods, Table~\ref{tab:main_results} reports representative high-performance operating
points corresponding to $T=1$ for fixed refresh and $\lambda_B=0.02$ for Agentic LinkDreamer; the complete communication--perception trade-off is analyzed in Section~\ref{sec:comm_tradeoff}.

\begin{table}[t]
\centering
\caption{Outage-frame cooperative-perception performance on the V2X-Real
V2V test split. Higher AP and recovery are better.}
\label{tab:main_results}

\footnotesize
\setlength{\tabcolsep}{3.5pt}
\renewcommand{\arraystretch}{1.05}

\begin{tabular}{lccc}
\hline
Method
& AP@0.5
& AP@0.7
& Recovery@0.5 (\%)
\\
\hline

Cooperative upper reference
& 91.60
& 77.44
& 100.0
\\

Ego-only
& 83.40
& 65.59
& --
\\

Degraded cooperation
& 82.50
& 68.43
& 0.0
\\

Relative-motion warp
& 85.10
& 70.80
& 28.6
\\

Always-imagine
& 86.45
& 62.70
& 43.4
\\

Fixed refresh
& 89.10
& 72.47
& 72.5
\\

\textbf{Agentic LinkDreamer}
& \textbf{90.95}
& \textbf{73.31}
& \textbf{92.9}
\\
\hline
\end{tabular}
\end{table}

Relative-motion warp improves AP@0.5 from 82.50 to 85.10 by reusing the most recently received collaborator representation after motion compensation, showing that temporally aligned historical information can partially mitigate communication outages without generative prediction. Always-imagine further increases AP@0.5 to 86.45 by reconstructing the missing collaborator representation with the latent world model. However, its AP@0.7 decreases to
62.70, indicating that unconditional use of imagined features is less reliable under the stricter localization criterion. This behavior motivates explicitly estimating prediction reliability before injecting imagined collaborator features into the detector. Fixed refresh achieves an AP@0.5 of 89.10 and recovers 72.5\% of the cooperative-perception gap, confirming that periodically reacquiring current
collaborator information can substantially mitigate communication-induced degradation. Agentic LinkDreamer further improves AP@0.5 to 90.95 and recovers 92.9\% of the gap, remaining only 0.65 AP below the
cooperative upper reference of 91.60.

Unlike always-imagine, Agentic LinkDreamer can reject unreliable predictions and select among ego-only perception, historical feature reuse, latent imagination, and feasible refresh actions according to the current decision
state. The comparison with always-imagine is interpreted jointly with the ablation study in Section~\ref{sec:ablation}, where the contributions of
adaptive decision making, detector adaptation, predictive uncertainty, and relative-motion conditioning are examined separately.

Fixed refresh additionally serves as a communication-aware reference. Rather than updating collaborator information at a predefined interval, Agentic LinkDreamer adapts its communication and recovery behavior according
to the geometry-derived link-quality proxy, memory age, predictive uncertainty, and expected perceptual benefit. The corresponding communication overhead and perception accuracy are jointly evaluated in Section~\ref{sec:comm_tradeoff}.

\subsection{Robustness to Outage Duration}
\label{sec:outage_duration}

We further evaluate robustness as the missing-collaborator interval becomes longer. Outage frames are divided according to the consecutive outage depth $\tau_t$ into three groups:
$\mathrm{short}:~\tau_t=1\text{--}2$,
$\mathrm{medium}:~\tau_t=3\text{--}6$, and
$\mathrm{long}:~\tau_t=7\text{--}10$.
Short outages mainly reflect temporal continuity and motion compensation, whereas longer interruptions introduce greater scene evolution, accumulated relative-motion error, and uncertainty in the missing collaborator view. Table~\ref{tab:outage_duration} reports AP@0.5 over the three outage-depth ranges.

\begin{table}[t]
\centering
\caption{Outage-frame AP@0.5 under different outage depths.}
\label{tab:outage_duration}
\small
\begin{tabular}{lccc}
\hline
Method
& $\tau_t=1$--$2$
& $\tau_t=3$--$6$
& $\tau_t=7$--$10$
\\
\hline
Degraded cooperation
& 83.10 & 82.50 & 78.20
\\
Relative-motion warp
& 86.50 & 83.20 & 80.30
\\
Always-imagine
& 87.30 & 85.70 & 82.80
\\
\textbf{Agentic LinkDreamer}
& \textbf{91.20}
& \textbf{90.10}
& \textbf{89.40}
\\
\hline
\end{tabular}
\end{table}

The results show that historical-feature reuse becomes increasingly sensitive to prolonged communication outages. Relative-motion warp decreases from 86.50 AP for short outages to 80.30 AP for long outages, corresponding to a
6.20-point degradation. Always-imagine is more robust, decreasing from 87.30 to 82.80 AP, which indicates that latent prediction can better preserve collaborator information as the most recently received feature becomes stale.

Agentic LinkDreamer exhibits substantially stronger robustness across all outage ranges. Its AP@0.5 decreases only from 91.20 for short outages to 89.40 for long outages, a reduction of 1.80 points. Moreover, its advantage
over always-imagine increases from 3.90 AP at $\tau_t=1$--$2$ to 6.60 AP at $\tau_t=7$--$10$. This widening gap indicates that adaptive action selection becomes increasingly important as recursive prediction becomes less reliable during prolonged interruptions. The improved long-outage performance is consistent with the controller's use of outage depth, memory age, and predictive uncertainty. Rather than relying
on latent imagination throughout the entire interruption, Agentic LinkDreamer can switch among imagined features, historical feature reuse, ego-only fallback, and feasible communication refresh according to the estimated
reliability of the current collaborator representation. The corresponding action distributions are analyzed together with the neuro-fuzzy policy behavior in Section~\ref{sec:ablation}.

\subsection{Communication--Perception Trade-off}
\label{sec:comm_tradeoff}

The multi-objective formulation in Section~3 balances perception quality against communication overhead. We vary the communication penalty $\lambda_B$ while keeping the remaining decision parameters fixed and evaluate the resulting operating points in terms of outage-frame AP and
average transmitted data volume. A larger $\lambda_B$ imposes a stronger penalty on communication and therefore encourages greater reliance on causal memory and latent imagination, whereas a smaller $\lambda_B$ permits more frequent collaborator refreshes.

\begin{figure}[t]
\centering
\includegraphics[width=0.88\linewidth]{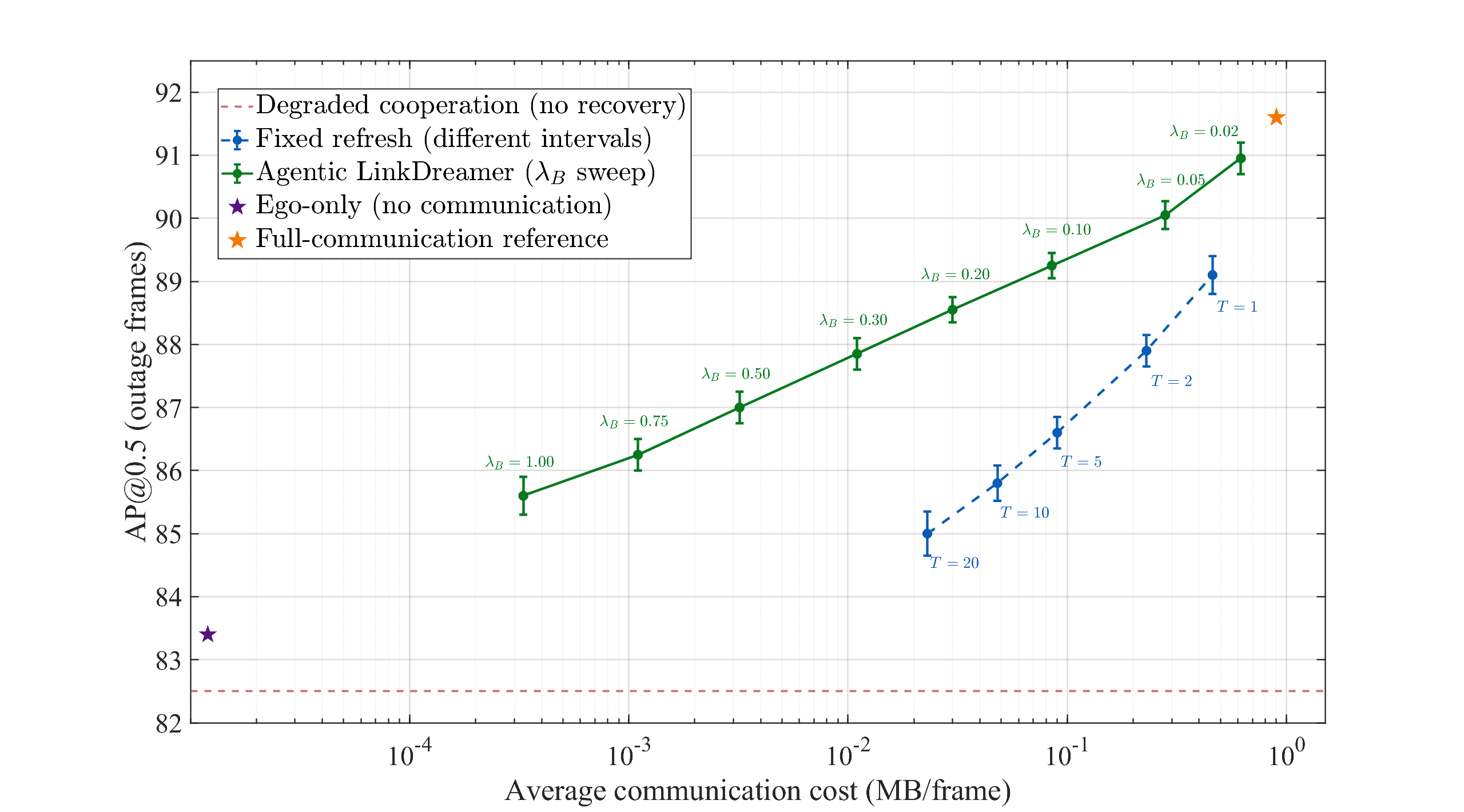}
\caption{Communication--perception trade-off on the V2X-Real V2V test split. Agentic LinkDreamer varies the communication penalty $\lambda_B$, while fixed-refresh baselines use different refresh intervals $T$. Error bars denote variation across independent runs.}
\label{fig:comm_pareto}
\end{figure}

Fixed-refresh baselines are evaluated using different refresh intervals $T$ so that their operating points span communication budgets comparable to those of Agentic LinkDreamer. The resulting communication--perception trade-off is shown in Fig.~\ref{fig:comm_pareto}. As $\lambda_B$ decreases, Agentic LinkDreamer progressively increases its
use of collaborator refresh actions and moves toward the cooperative upper reference. The resulting operating points exhibit a smooth transition from communication-efficient prediction to communication-intensive cooperative
perception, demonstrating that the proposed controller can adapt its operating regime through the communication penalty.

Agentic LinkDreamer consistently occupies a more favorable
communication--perception region than fixed periodic refresh. For example, at approximately $0.09$~MB/frame, the operating point with $\lambda_B=0.10$ achieves about 89.25 AP@0.5, compared with 86.60 for fixed refresh with $T=5$, yielding an improvement of approximately 2.65 AP at a comparable communication cost. At a lower communication budget, $\lambda_B=0.30$ achieves about 87.85 AP using approximately
0.011~MB/frame, while the $T=20$ fixed-refresh baseline achieves about 85.00 AP with approximately 0.023~MB/frame.

At the high-performance operating point with $\lambda_B=0.02$, Agentic Link-Dreamer reaches 90.95 AP@0.5 with an average communication cost of approximately 0.62~MB/frame. This is only 0.65 AP below the cooperative
upper reference of 91.60 while reducing the transmitted data volume by approximately 31\% relative to full communication at 0.90~MB/frame. These results indicate that the proposed adaptive controller retains most of the cooperative-perception benefit without requiring continuous
high-rate transmission.

The observed trade-off is consistent with the decision mechanism of Agentic LinkDreamer. Large values of $\lambda_B$ favor memory reuse and latent imagination, whereas smaller values permit more frequent low-rate or high-rate refresh actions when the expected perceptual benefit justifies the
additional communication cost. The corresponding action-selection behavior is further analyzed in Section~\ref{sec:ablation}.

\subsection{Ablation and Neuro-Fuzzy Decision Analysis}
\label{sec:ablation}

We further investigate the contribution of the principal components of Agentic LinkDreamer through the ablation study in Table~\ref{tab:ablation}. The variants separately examine adaptive action selection, the neuro-fuzzy decision structure, predictive uncertainty, relative-motion conditioning, and task-level detector adaptation.

\begin{table}[t]
\centering
\caption{Ablation study on communication-outage frames.}
\label{tab:ablation}

\footnotesize
\setlength{\tabcolsep}{5pt}
\renewcommand{\arraystretch}{1.08}

\begin{tabular}{lccc}
\hline
Variant
& AP@0.5
& AP@0.7
& Recovery@0.5 (\%)
\\
\hline

\textbf{Full Agentic LinkDreamer}
& \textbf{90.95}
& \textbf{73.31}
& \textbf{92.9}
\\

Always-imagine / no policy
& 86.45
& 62.70
& 43.4
\\

MLP decision policy
& 89.60
& 71.85
& 78.0
\\

w/o uncertainty input
& 89.25
& 70.95
& 74.2
\\

w/o relative-motion conditioning
& 88.70
& 70.20
& 68.1
\\

w/o detector adaptation
& 84.45
& 58.70
& 21.4
\\

\hline
\end{tabular}
\end{table}

The full Agentic LinkDreamer achieves 90.95 AP@0.5 and a recovery ratio of 92.9\%, providing the strongest performance among all variants. Removing adaptive action selection and always injecting the imagined collaborator representation reduces AP@0.5 to 86.45 and AP@0.7 to 62.70. The corresponding
4.50-point AP@0.5 degradation indicates that latent prediction alone is insufficient when prediction reliability varies across outage conditions. Adaptive selection of imagination, historical reuse, ego-only fallback, and
communication refresh therefore contributes substantially to robust recovery.

Replacing the Takagi--Sugeno neuro-fuzzy controller with an MLP policy reduces AP@0.5 from 90.95 to 89.60 and AP@0.7 from 73.31 to 71.85. Since both controllers use the same decision-state inputs, admissible-action mask, and training objective, this result indicates that the proposed neuro-fuzzy
structure provides an additional benefit beyond generic nonlinear action selection while retaining an interpretable decision representation. Predictive uncertainty also has a measurable contribution. Removing the uncertainty input decreases AP@0.5 to 89.25 and recovery to 74.2\%. Without an explicit estimate of prediction reliability, the controller is
less able to identify when recursively imagined collaborator features should be rejected or replaced by safer fallback or refresh actions. Relative-motion conditioning produces a further performance gain. Removing the relative-motion input reduces AP@0.5 to 88.70 and AP@0.7 to 70.20, corresponding to a 2.25-point AP@0.5 drop relative to the full model. This
confirms that inter-agent motion information is important for maintaining temporally consistent collaborator representations as the ego--collaborator geometry evolves during an outage.

The largest degradation is observed without detector adaptation. Keeping the cooperative fusion module and detection head frozen reduces AP@0.5 to 84.45
and AP@0.7 to 58.70, with only 21.4\% recovery of the cooperative-perception gap. This result indicates a substantial distribution mismatch between real
collaborator features and world-model-generated representations and confirms the importance of adapting the downstream perception modules to both feature
sources.

\begin{figure}[t]
\centering

\begin{subfigure}[t]{0.32\linewidth}
    \centering
    \includegraphics[width=\linewidth]{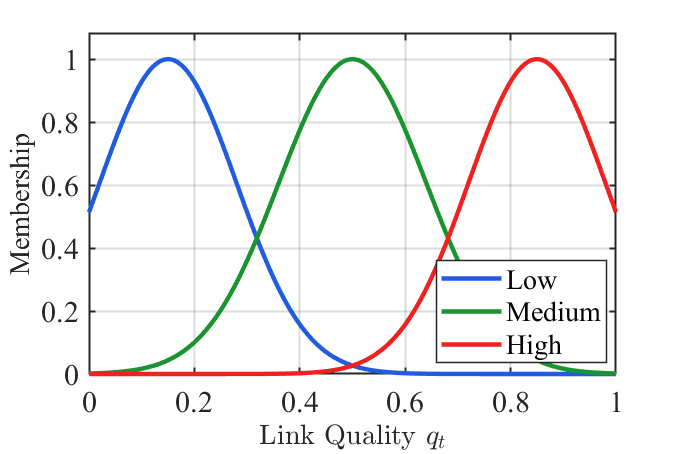}
    \caption{Link quality $q_t$.}
    \label{fig:fuzzy_link}
\end{subfigure}
\hfill
\begin{subfigure}[t]{0.32\linewidth}
    \centering
    \includegraphics[width=\linewidth]{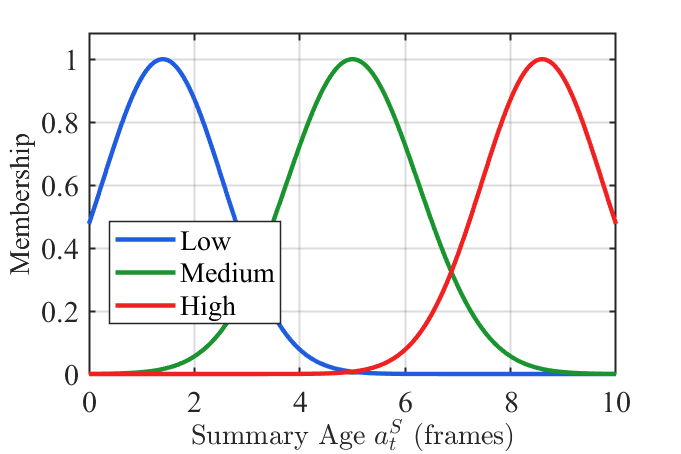}
    \caption{Summary age $a_t^{S}$.}
    \label{fig:fuzzy_age}
\end{subfigure}
\hfill
\begin{subfigure}[t]{0.32\linewidth}
    \centering
    \includegraphics[width=\linewidth]{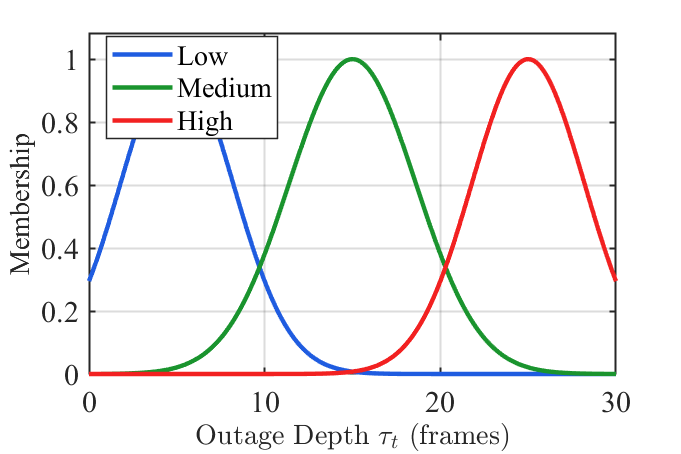}
    \caption{Outage depth $\tau_t$.}
    \label{fig:fuzzy_outage}
\end{subfigure}

\vspace{1.5mm}

\begin{subfigure}[t]{0.32\linewidth}
    \centering
    \includegraphics[width=\linewidth]{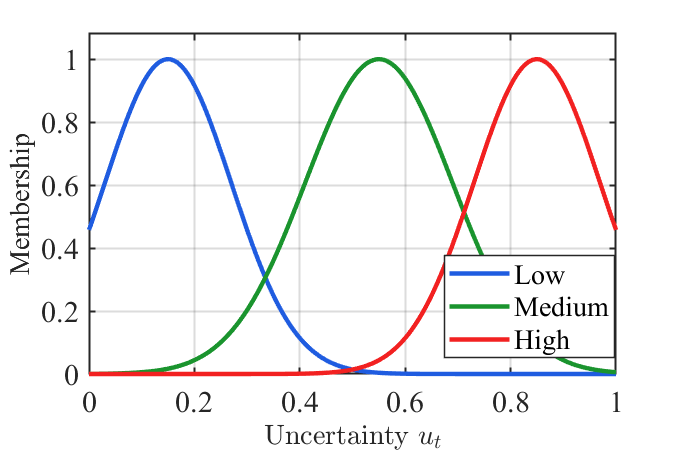}
    \caption{Predictive uncertainty $u_t$.}
    \label{fig:fuzzy_uncertainty}
\end{subfigure}
\hspace{5mm}
\begin{subfigure}[t]{0.32\linewidth}
    \centering
    \includegraphics[width=\linewidth]{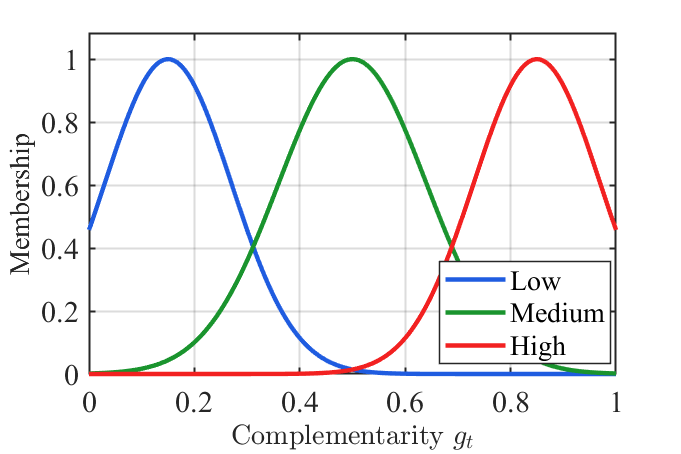}
    \caption{Perceptual complementarity $g_t$.}
    \label{fig:fuzzy_complementarity}
\end{subfigure}

\vspace{1.5mm}

\begin{subfigure}[t]{0.43\linewidth}
    \centering
    \includegraphics[width=\linewidth]{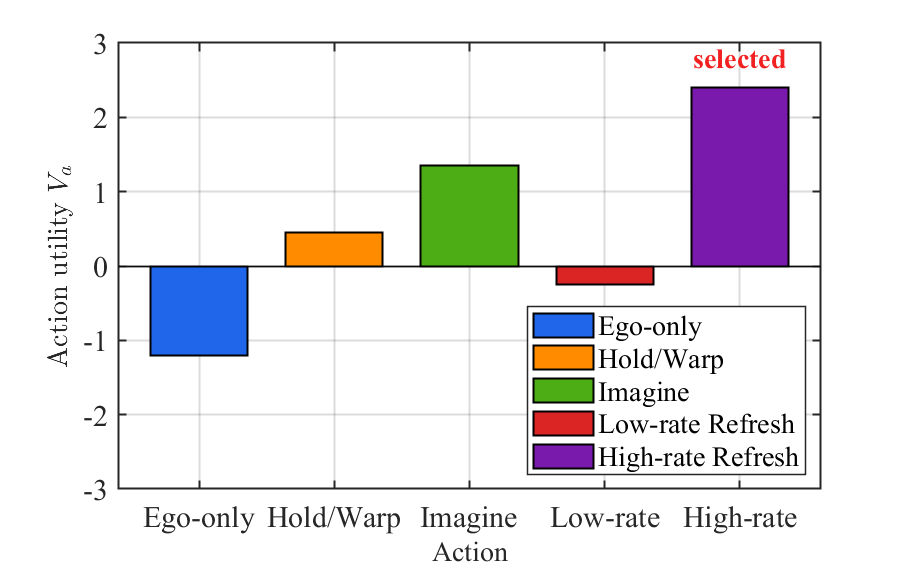}
    \caption{Action utilities for an example state.}
    \label{fig:fuzzy_action}
\end{subfigure}
\hfill
\begin{subfigure}[t]{0.54\linewidth}
    \centering
    \includegraphics[width=\linewidth]{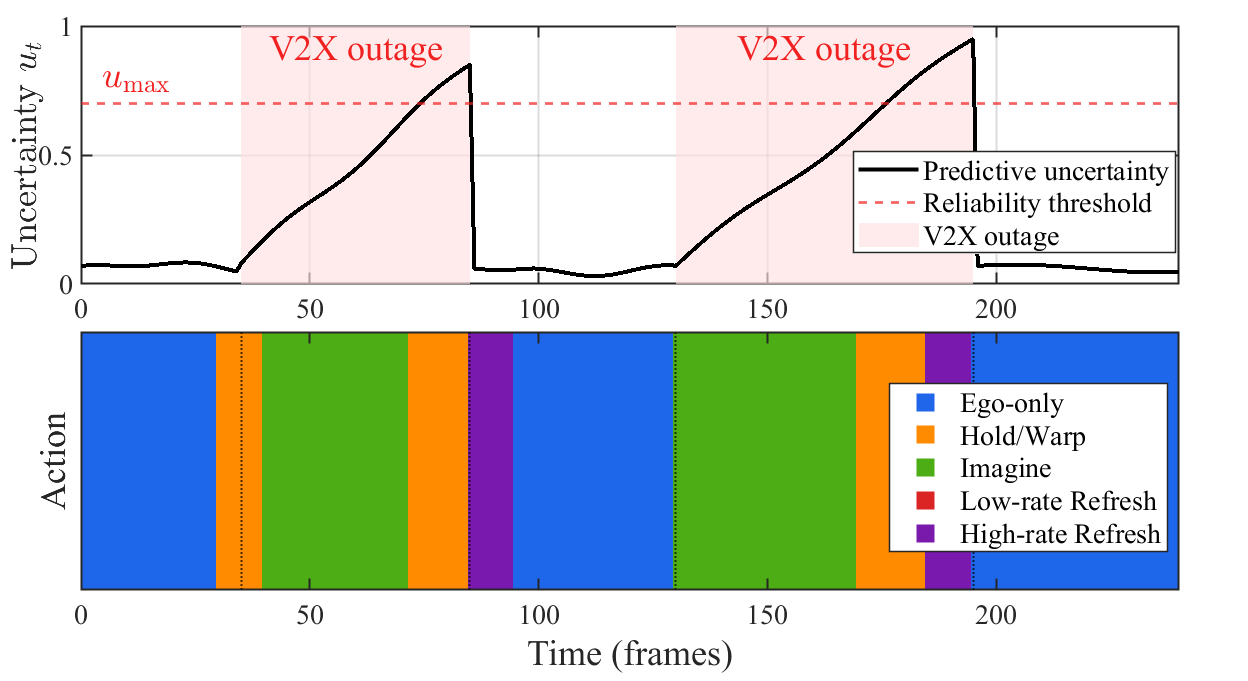}
    \caption{Decision trajectory during V2X outages.}
    \label{fig:fuzzy_trajectory}
\end{subfigure}

\caption{Neuro-fuzzy controller interpretability analysis.
(a)--(e) Learned membership functions; (f) action utilities for an example state; and (g) uncertainty-aware action selection during V2X outages.}
\label{fig:fuzzy_analysis}
\end{figure}

To further examine the interpretability of the proposed controller, Fig.~\ref{fig:fuzzy_analysis} visualizes the learned fuzzy partitions, action utilities, and a representative uncertainty-aware decision trajectory.

Figs.~\ref{fig:fuzzy_link}--\ref{fig:fuzzy_complementarity} show the learned low, medium, and high fuzzy regions for link quality $q_t$, summary age $a_t^{S}$, outage depth $\tau_t$, predictive uncertainty $u_t$, and perceptual complementarity $g_t$. The overlapping membership functions provide smooth transitions between neighboring operating regions instead of
hard state thresholds, which enables gradual changes in action preference as the communication and perception conditions evolve.

Fig.~\ref{fig:fuzzy_action} shows the action utilities produced for a representative decision state. Among the five admissible actions, high-rate refresh receives the highest utility, whereas ego-only perception and low-rate refresh receive negative utilities. This example illustrates how the
controller evaluates recovery quality and communication cost jointly rather than following a predetermined refresh schedule. The temporal behavior in Fig.~\ref{fig:fuzzy_trajectory} further demonstrates the role of predictive uncertainty during intermittent communication. Within each shaded V2X outage interval, $u_t$ progressively increases as the
missing-collaborator duration becomes longer. The controller initially relies on latent imagination while prediction uncertainty remains moderate, then moves toward historical feature reuse as uncertainty approaches the reliability threshold $u_{\max}$. High-rate refresh is selected around link recovery to reacquire an up-to-date collaborator representation, after which the predictive uncertainty rapidly decreases. 
Together, the quantitative ablations and the policy visualization show that the performance gain of Agentic LinkDreamer does not arise from latent prediction alone. Relative-motion-conditioned world modeling, explicit
uncertainty estimation, task-level detector adaptation, and interpretable neuro-fuzzy action selection jointly enable the controller to switch between prediction, fallback, and communication according to the reliability and expected value of collaborator information.

\subsection{Qualitative Analysis}
\label{sec:qualitative}

\begin{figure}[t]
\centering
\includegraphics[width=0.98\linewidth]{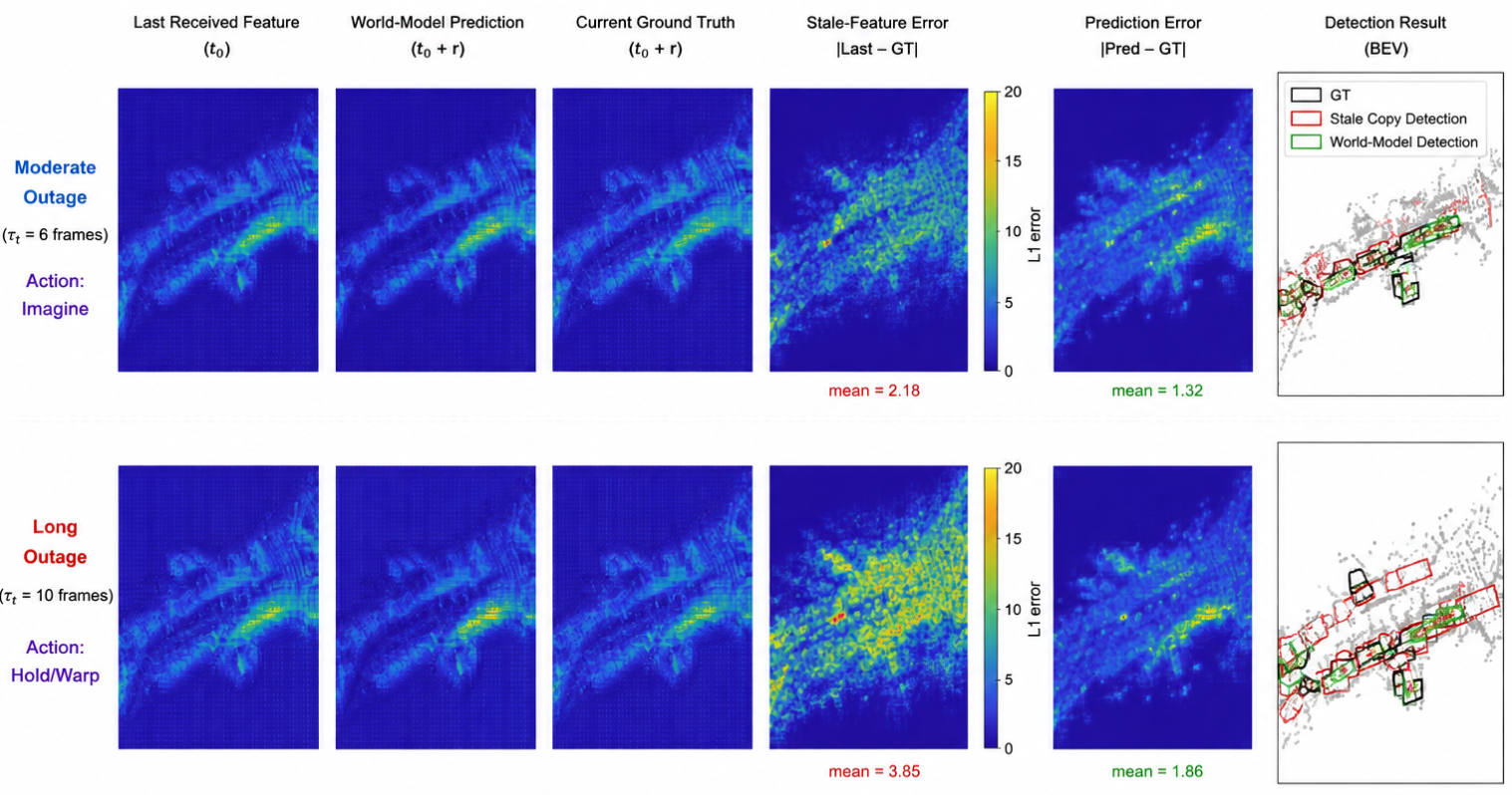}
\caption{Qualitative analysis during V2X communication interruption. Representative moderate- and long-outage frames illustrate the latest received collaborator feature, world-model prediction, offline ground-truth feature, feature errors, selected action, and final BEV detection result.}
\label{fig:qualitative_results}
\end{figure}

Finally, we qualitatively examine representative V2X-outage frames to illustrate the interaction between latent prediction and adaptive action selection. Fig.~\ref{fig:qualitative_results} shows the latest received
collaborator feature, the world-model prediction, the corresponding real collaborator feature used only for offline evaluation, the feature-error maps, the selected action, and the final BEV detection result.

For the moderate outage, the world model provides a more temporally consistent collaborator representation than direct reuse of the stale feature, and the controller selects latent imagination. As the outage becomes longer, prediction uncertainty and accumulated temporal error increase, causing the controller to adopt a more conservative historical-feature reuse action. This behavior illustrates that Agentic LinkDreamer does not rely on latent prediction unconditionally, but adaptively selects the collaborator information source according to prediction reliability and the current decision state. The ground-truth collaborator feature and error maps in Fig.~\ref{fig:qualitative_results} are used only for offline analysis and are not available to the controller during an outage.
\section{Conclusion}
\label{sec:conclusion}

This paper formulates cooperative perception under intermittent V2X communication as an adaptive communication--perception decision problem. \method{} integrates causal collaborator memory, a relative-motion-conditioned
latent world model, uncertainty estimation, and an interpretable neuro-fuzzy controller to adaptively select among ego-only perception, historical-feature
reuse, latent imagination, and communication refresh. The resulting framework balances perception accuracy, communication overhead, latency, prediction
reliability, and action stability within a unified decision process. Experiments on the V2X-Real V2V split demonstrate that \method{} achieves 90.95 AP@0.5 on outage frames and recovers 92.9\% of the cooperative-perception performance gap. At the high-performance operating point, it remains only
0.65 AP below the cooperative upper reference while reducing communication volume by approximately 31\% relative to continuous high-rate cooperation. The ablation and qualitative results further show that relative-motion-aware
prediction, uncertainty estimation, detector adaptation, and neuro-fuzzy action selection jointly contribute to robust performance under prolonged communication interruptions. Future work will extend the framework to multiple heterogeneous collaborators and evaluate it with measured V2X communication traces.


\section*{Acknowledgement}
The authors are grateful to King Saud University, Riyadh, Saudi Arabia for funding this work through Ongoing Research Funding Program under Grant ORF-2026-1007.

\section*{Declaration of competing interest}
The authors declare that they have no known competing financial interests or personal relationships that could have appeared to influence the work reported in this paper.

\section*{Declaration of generative AI and AI-assisted technologies
in the writing process}
During the preparation of this work, the authors used ChatGPT (OpenAI) to assist with language refinement and improve the
readability of the manuscript. The authors reviewed and edited the output as needed and take full responsibility for the content
of the published article.

\appendix

\bibliographystyle{elsarticle-num}
\bibliography{references}

\end{document}